\documentclass[structabstract]{aa}  
\usepackage{graphicx}
\usepackage{txfonts}
\usepackage{natbib}
\newcommand{\um}{\,$\mu$m}
\newcommand{\oi}{[O\,{\sc i}]}
\newcommand{\cii}{[C\,{\sc ii}]}
\newcommand{\ci}{[C\,{\sc i}]}

\newcommand{\neii}{[Ne\,{\sc ii}]}

\newcommand{\suiii}{[S\,{\sc iii}]}
\newcommand{\piii}{[P\,{\sc iii}]}
\newcommand{\ariii}{[Ar\,{\sc iii}]}
\newcommand{\hii}{H\,{\sc ii}}
\newcommand{\pahn}{PAH$^0$}
\newcommand{\pahp}{PAH$^+$}
\newcommand{\pahx}{PAH$^x$}
\newcommand{\hh}{H$_2$}
\newcommand{\nh}{$n_{\mathrm{H}}$}
\newcommand{\cc}{cm$^{-3}$}
\newcommand{\pe}{$\epsilon_\mathrm{pe}$}
\newcommand{\pepah}{$\epsilon_\mathrm{PAH}$}
\bibpunct{(}{)}{;}{a}{}{,}
\begin{document}
   \title{Probing the role of polycyclic aromatic hydrocarbons in the photoelectric heating within photodissociation regions\thanks{\textit{Herschel} is an ESA space observatory with science instruments provided by European-led Principal Investigator consortia and with important participation from NASA. This work is based in part on observations made with the Spitzer Space Telescope, which is operated by the Jet Propulsion Laboratory, California Institute of Technology under a contract with NASA.}}

   \subtitle{}

   \author{Yoko Okada \inst{1}
          \and
          Paolo Pilleri \inst{2,3,4}
          \and
          Olivier Bern\'{e} \inst{5,6}
          \and
          Volker Ossenkopf \inst{1}
          \and
          Asunci\'{o}n Fuente \inst{3}
          \and
          Javier R. Goicoechea \inst{2}
          \and
          Christine Joblin \inst{5,6}
          \and
          Carsten Kramer \inst{7}
          \and
          Markus R\"{o}llig \inst{1}
          \and
          David Teyssier \inst{8}
          \and
          Floris F. S. van der Tak \inst{9,10}
          }

   \institute{I. Physikalisches Institut der Universit\"{a}t zu K\"{o}ln, Z\"{u}lpicher Stra{\ss}e 77, 50937 K\"{o}ln, Germany
              \email{okada@ph1.uni-koeln.de}
         \and
             Centro de Astrobiolog\'ia, CSIC-INTA, 28850, Madrid, Spain	
         \and
             Observatorio Astron\'omico Nacional (OAN), Apdo. 112, 28803 Alcal\'a de Henares (Madrid), Spain	
	 \and
             Los Alamos National Laboratory, Los Alamos, NM 87545, USA. 
         \and
             Universit\'e de Toulouse; UPS-OMP; IRAP;  Toulouse, France
         \and
             CNRS; IRAP; 9 Av. colonel Roche, BP 44346, F-31028 Toulouse cedex 4, France
         \and
             Instituto de Radioastronom\'{i}a Milim\'{e}trica, Av. Divina Pastora 7, Nucleo Central, 18012 Granada, Spain
         \and
             European Space Astronomy Centre, ESA, PO Box 78, 28691, Villanueva de la Ca\~{n}ada, Madrid, Spain
         \and
             SRON Netherlands Institute for Space Research, P.O. Box 800, 9700 AV Groningen, The Netherlands 
         \and
             Kapteyn Astronomical Institute, University of Groningen, PO Box 800, 9700 AV Groningen, The Netherlands 
}

   \date{Received; accepted}

 
  \abstract
   {}
   {We observationally investigate the relation between the photoelectric heating efficiency in photodissociation regions (PDRs) and the charge of polycyclic aromatic hydrocarbons (PAHs), which are considered to play a key role in photoelectric heating.}
   {Using PACS onboard \textit{Herschel}, we observed six PDRs spanning a wide range of far-ultraviolet radiation fields ($G_0=100$--$10^5$).  To measure the photoelectric heating efficiency, we obtained the intensities of the main cooling lines in these PDRs, i.e., the \oi\ 63\um, 145\um, and \cii\ 158\um, as well as the far-infrared (FIR) continuum intensity.  We used \textit{Spitzer}/IRS spectroscopic mapping observations to investigate the mid-infrared (MIR; 5.5--14\um) PAH features in the same regions.  We decomposed the MIR PAH emission into that of neutral (\pahn) and positively ionized (\pahp) species to derive the fraction of the positively charged PAHs in each region, and compare it to the photoelectric heating efficiency.}
   {The heating efficiency traced by (\oi\ 63\um\ $+$ \oi\ 145\um\ $+$ \cii\ 158\um) / TIR, where TIR is the total infrared flux, ranges between 0.1\% and 0.9\% in different sources, and the fraction of \pahp\ relative to (\pahn $+$ \pahp) spans from 0 ($+11$)\% to 87 ($\pm 10$)\%.  All positions with a high \pahp\ fraction show a low heating efficiency, and all positions with a high heating efficiency have a low \pahp\ fraction, supporting the scenario in which a positive grain charge results in a decreased heating efficiency.  Theoretical estimates of the photoelectric heating efficiency show a stronger dependence on the charging parameter $\gamma=G_0 T^{1/2}/n_e$ than the observed efficiency reported in this study, and the discrepancy is significant at low $\gamma$.  The photoelectric heating efficiency on PAHs, traced by (\oi\ 63\um\ $+$ \oi\ 145\um\ $+$ \cii\ 158\um) / (PAH-band emission $+$ \oi\ 63\um\ $+$ \oi\ 145\um\ $+$ \cii\ 158\um), shows a much better match between the observations and the theoretical estimates.}
   {The good agreement of the photoelectric heating efficiency on PAHs with a theoretical model indicates the dominant contribution of PAHs to the photoelectric heating.  This study demonstrates the fundamental role that PAHs have in photoelectric heating.  More studies of their charging behavior are crucial to understand the thermal balance of the interstellar medium.}


   \titlerunning{Probing the role of PAHs in the photoelectric heating within PDRs}
   \authorrunning{Y. Okada et al.}

   \maketitle

%

\section{Introduction}\label{sect:intro}

Photoelectric heating is a major heating process in photodissociation regions (PDRs), and its efficiency (\pe) is one of the key parameters to understanding the energy balance there.  Theoretical investigations suggest that small grains, in particular polycyclic aromatic hydrocarbons (PAHs), play a dominant role in photoelectric heating \citep{Bakes1994}.  \pe\ is defined as the fraction of energy absorbed by dust that is converted into kinetic energy of the ejected electrons and therefore into gas heating.  Observationally it can be estimated by measuring the ratio of the energy emitted in the gas cooling lines (\oi\ 63\um\ and \cii\ 158\um\ are the strongest ones from $Av\lesssim 5$ in PDRs) against the total UV energy absorbed by dust grains, which is probed by the dust infrared (IR) emission.  Previous observations show a wide variation of \pe\ in different sources, ranging from $10^{-4}$ in W49N, which is illuminated by an intense UV field \citep{Vastel2001} to 1--2\% in the relatively low-UV excited PDR of the Horsehead Nebula \citep{Goicoechea2009}, although mechanical heating also plays a role in W49 \citep{Nagy2012}.  \citet{Mizutani2004} showed a variation of \pe\ from 0.06 to 1.2\% across the $40^\prime \times 20^\prime$ area of the Carina Nebula, and an anti-correlation between \pe\ and the intensity of the local UV radiation field.  For extragalactic sources, \citet{Mookerjea2011} presented \pe\ of 0.3--1.2\% toward one \hii\ region in M\,33 at a resolution of 50\,pc.  These variations have been attributed to differences in the mean charge state of the grains, i.e., a positive grain charge results in a decreased efficiency, and the correlation with the intensity of the UV radiation field supports this interpretation.  However, considering two properties of different dust populations together, large grains that are the main carriers of the dust IR emission and small grains that are expected to be the main contributors of the photoelectric effect, limits the interpretation.  \citet{Habart2001} investigated the photoelectric heating efficiency on small grains and indicated that the observed spatial distribution of \oi\ 63\um\ and \cii\ 158\um\ in L1721 can be better explained by a model with a varying abundance of small grains across the cloud.  \citet{Joblin2010} concluded that the evolution of PAHs and very small grains at the border of PDRs should be considered to model the gas energetics.  Before \textit{Herschel}, estimating the gas-cooling energy in spatially resolved PDRs was only rarely possible because of the poor spatial resolution in the far-infrared (FIR) wavelength range.  To investigate the charge state of PAHs, mid-infrared (MIR) spectroscopic observations are needed, which can be provided by the Infrared Spectrograph \citep[IRS;][]{Houck2004} onboard the \textit{Spitzer Space Telescope} \citep{Werner2004}.  The most evident spectral difference between neutral PAHs ({\pahn}) and ionized PAHs ({\pahp}) is the strength of the 6--9\um\ complex relative to the 11.3\um\ feature, which is supported by laboratory experiments \citep{deFrees1993,Pauzat1994,Langhoff1996} and theoretical calculations \citep{Szczepanski1993a,Szczepanski1993b,Hudgins1995a,Hudgins1995b}.  Strong variations of the observed intensity of the individual bands have been seen in different environments, reflecting an evolution of the charge state of PAHs in different physical conditions \citep{Galliano2008,Sakon2004,Peeters2002,Joblin1996}.  Recently, \citet{Joblin2008} and \citet{Pilleri2012a} proposed an alternative method of deriving the fraction of \pahp\ using a spectral decomposition approach based on a few template spectra \citep{Berne2007,Rapacioli2005} that include \pahn, \pahp, and evaporating very small grains (eVSGs).

The combination of the observations by Photodetector Array Camera and Spectrometer \citep[PACS;][]{Poglitsch2010} onboard \textit{Herschel} \citep{Pilbratt2010} and the IRS onboard \textit{Spitzer} enables us to investigate the relation between \pe\ and the fraction of ionized PAHs in spatially resolved PDRs.  In this paper, we report the results for six PDRs studied in the WADI \citep[Warm And Dense Interstellar medium;][]{Ossenkopf2011} guaranteed time key program of \textit{Herschel}.  WADI is aimed to investigate the physics and chemistry of PDRs and shocked regions with a wide range of physical properties.  In this study, we investigate six PDRs located in five different regions that show clear detections of \cii\ and \oi\ with PACS and were observed with the IRS.

\section{Observations and data reduction}
\subsection{Targets} \label{subsec:targets}

\begin{table*}
\caption{General properties of the observed targets.  $G_0$ only gives typical values within the individual regions.  See Table~\ref{table:tir_and_uv} for local variations.}
\label{table:globalprop}
\centering
\begin{tabular}{lccccccc}
\hline
Object && Distance & \multicolumn{2}{c}{Exciting source} & $G_0^\textrm{a}$ & Reference\\
&& [kpc] & Name & Spectral Type & [Habing field] & \\
\hline\hline
Horsehead && 0.41 & $\sigma$ Ori & O9.5V & $100$ & 1,2 \\
Ced~201 && 0.42 & BD $+69^\circ$ 1231 & B9.5V & $300$ & 3\\
NGC~7023 & E & 0.43 & HD~200775 & B3Ve--B5 & $250$ & 4\\
& NW & 0.43 & HD~200775 & B3Ve--B5 & $2600$ & 4\\
Carina & N & 2.35 & Trumpler 14 & O3 & $7000$ & 5\\
Mon~R2 && 0.85 & IRS1 & B0 & $5\times 10^5$ & 6,7\\
\hline
\end{tabular}
\begin{list}{}{}
\item[$^\textrm{a}$] See text (Sect.~\ref{subsec:targets}) for the definition.
\item[Reference:] $^1$ \citet{Menten2007}, $^2$ \citet{Habart2005}, $^3$ \citet{YoungOwl2002}, $^4$ \citet{Pilleri2012a}, $^5$ \citet{Kramer2008}, $^6$ \citet{Downes1975}, $^7$ \citet{Rizzo2003}
\end{list}
\end{table*}

\begin{figure*}
\centering
\includegraphics[bb=72 170 540 612,width=0.45\textwidth,clip]{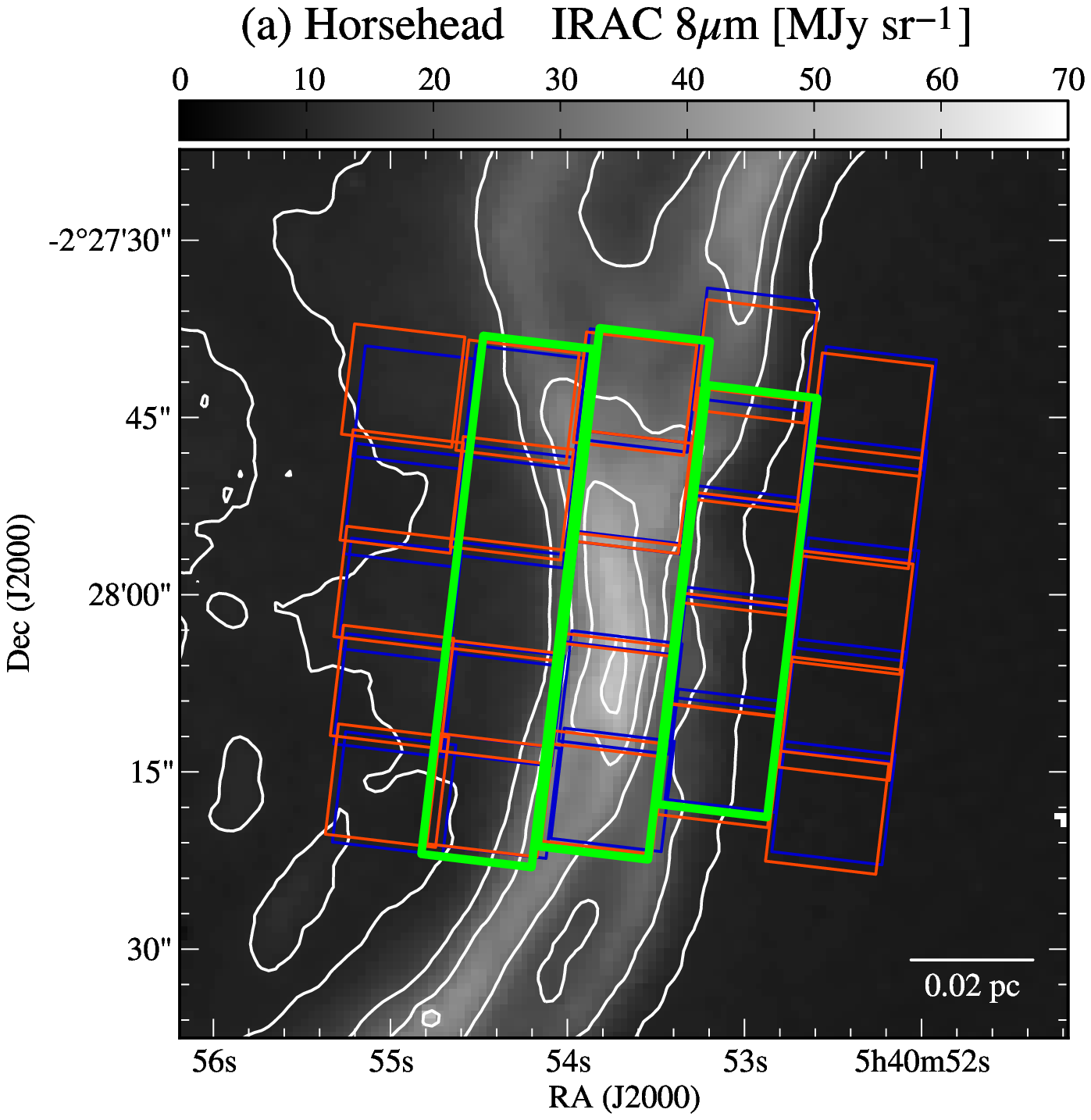}
\includegraphics[bb=72 170 540 612,width=0.45\textwidth,clip]{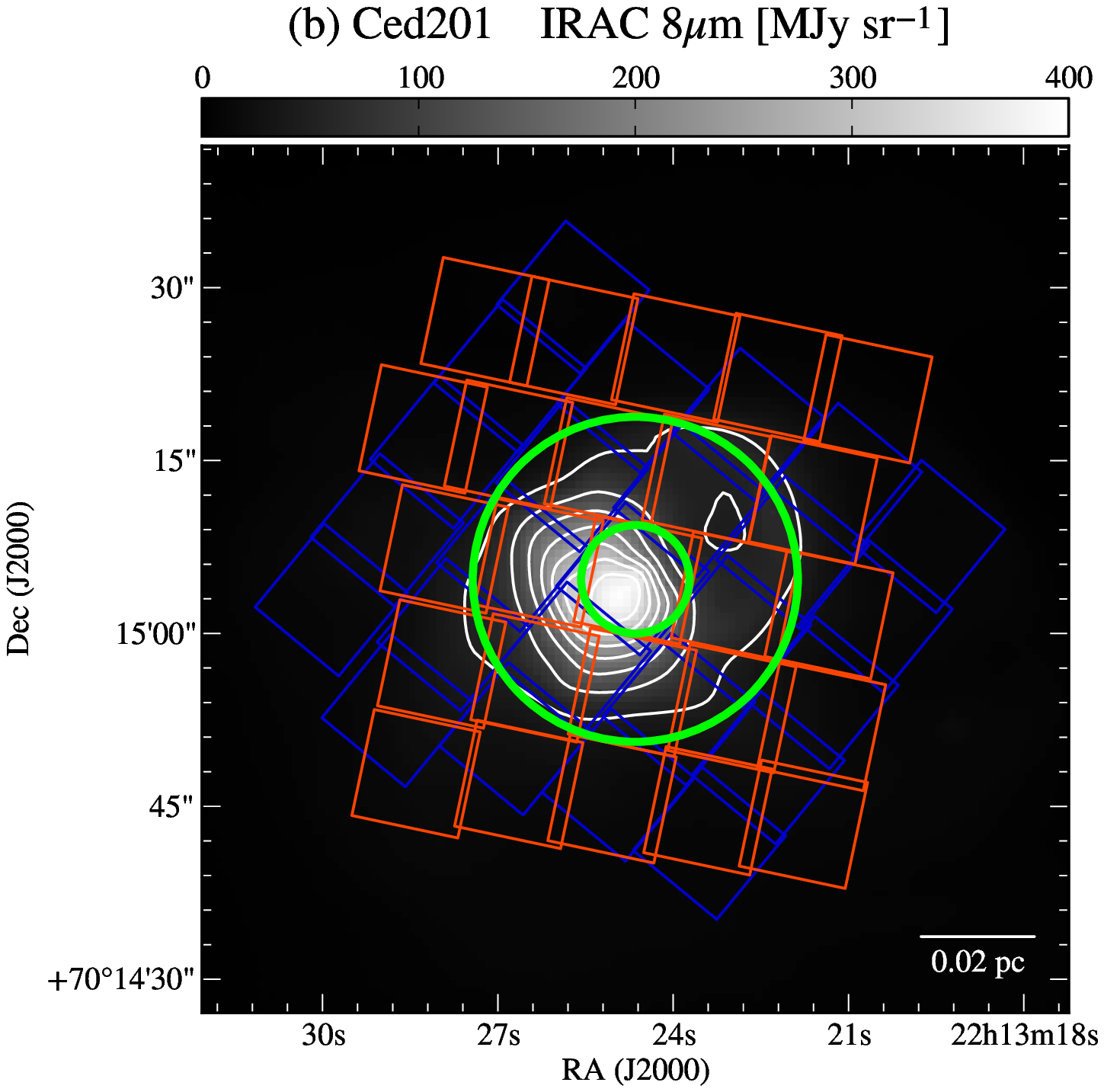}
\includegraphics[bb=72 170 540 612,width=0.45\textwidth,clip]{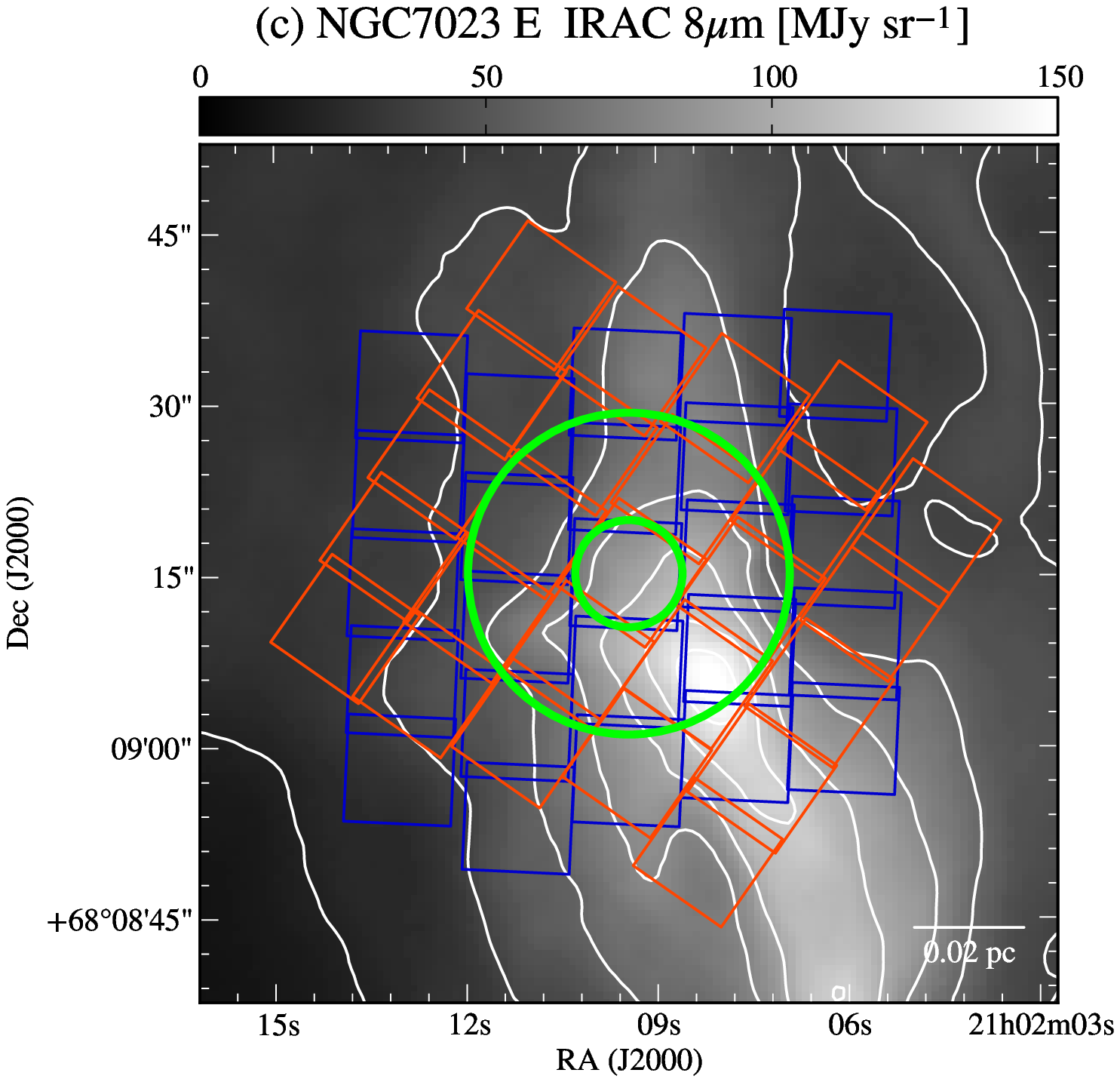}
\includegraphics[bb=72 170 540 612,width=0.45\textwidth,clip]{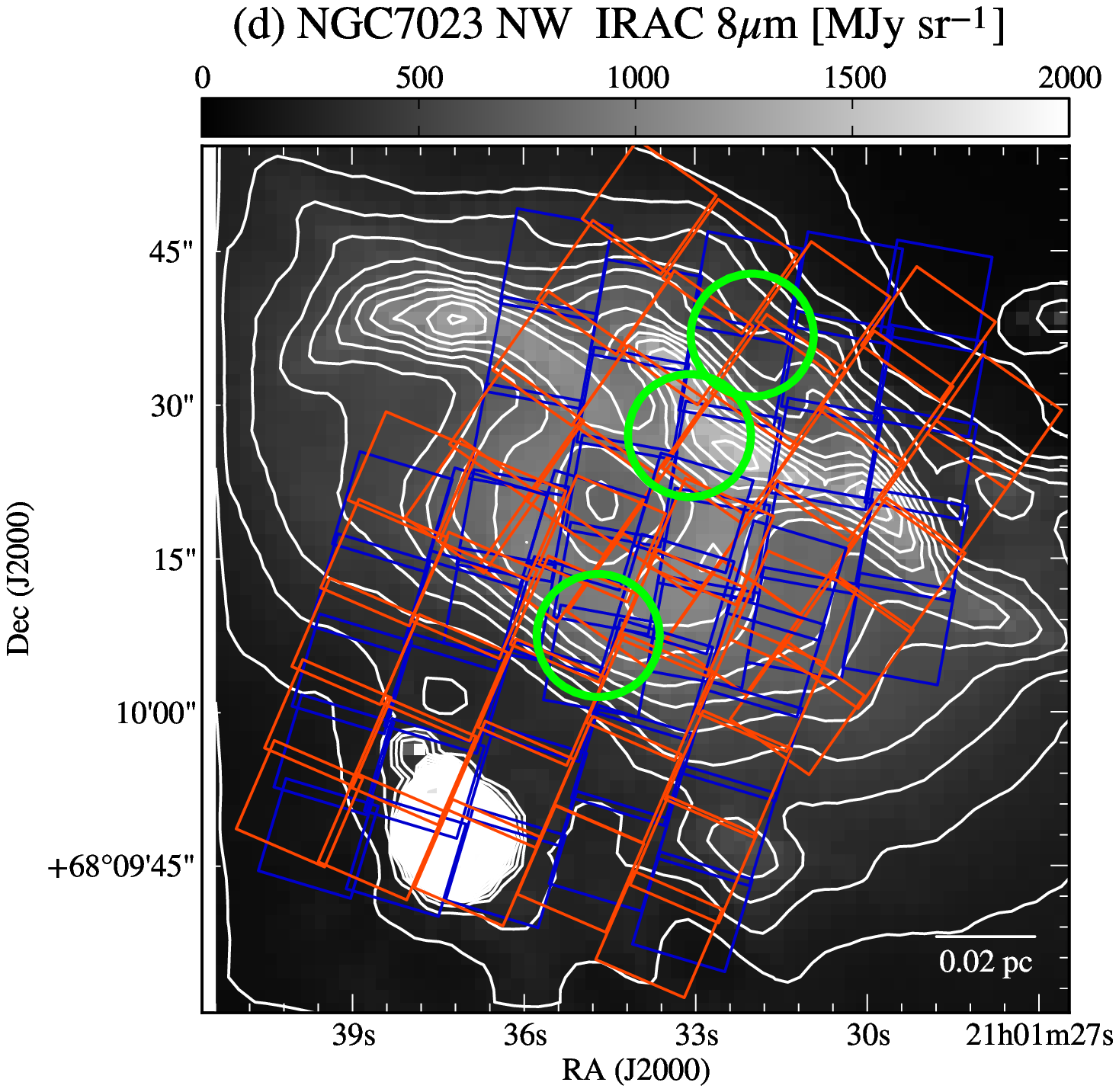}
\includegraphics[bb=72 170 540 612,width=0.45\textwidth,clip]{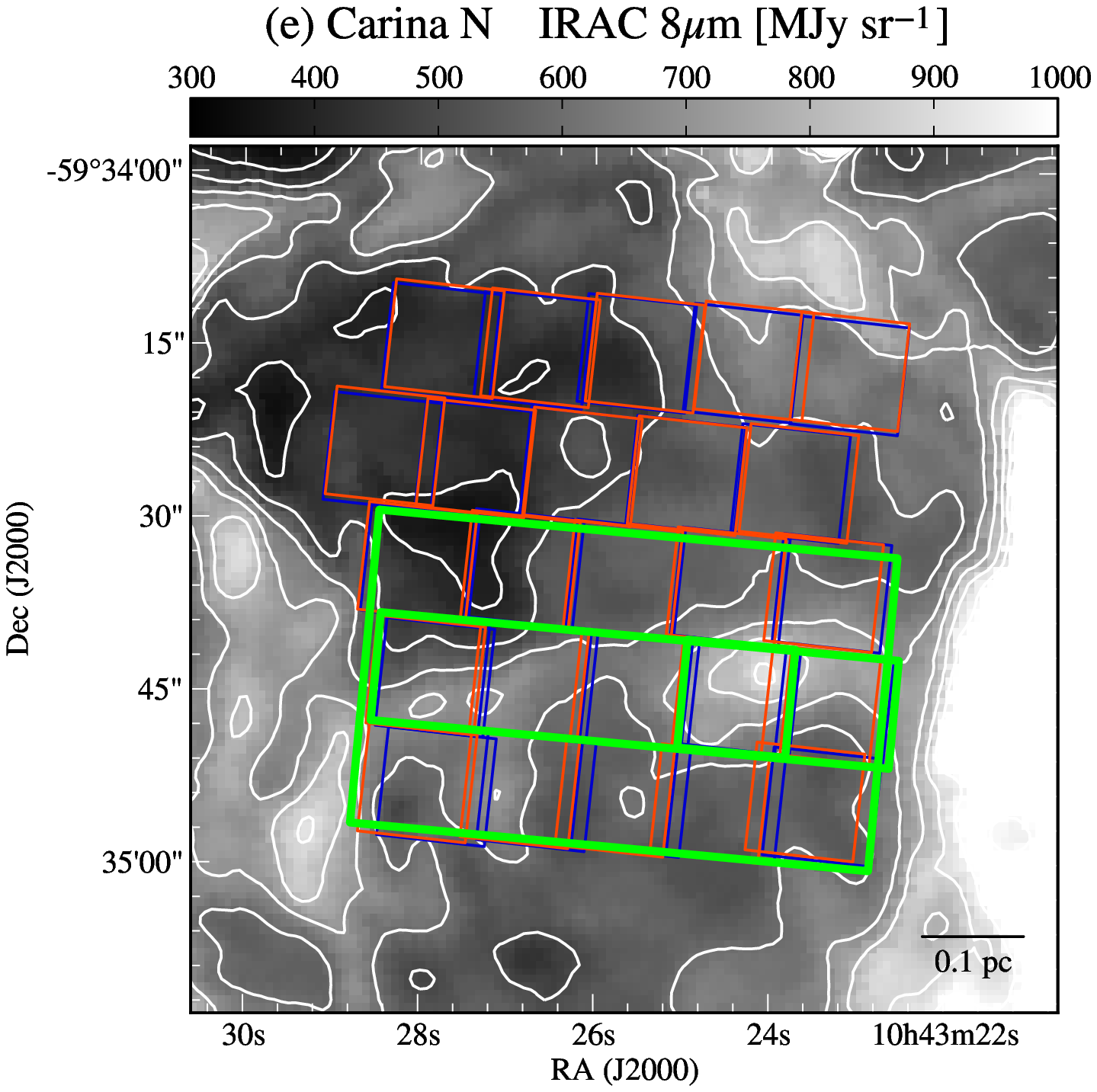}
\includegraphics[bb=54 170 558 612,width=0.45\textwidth,clip]{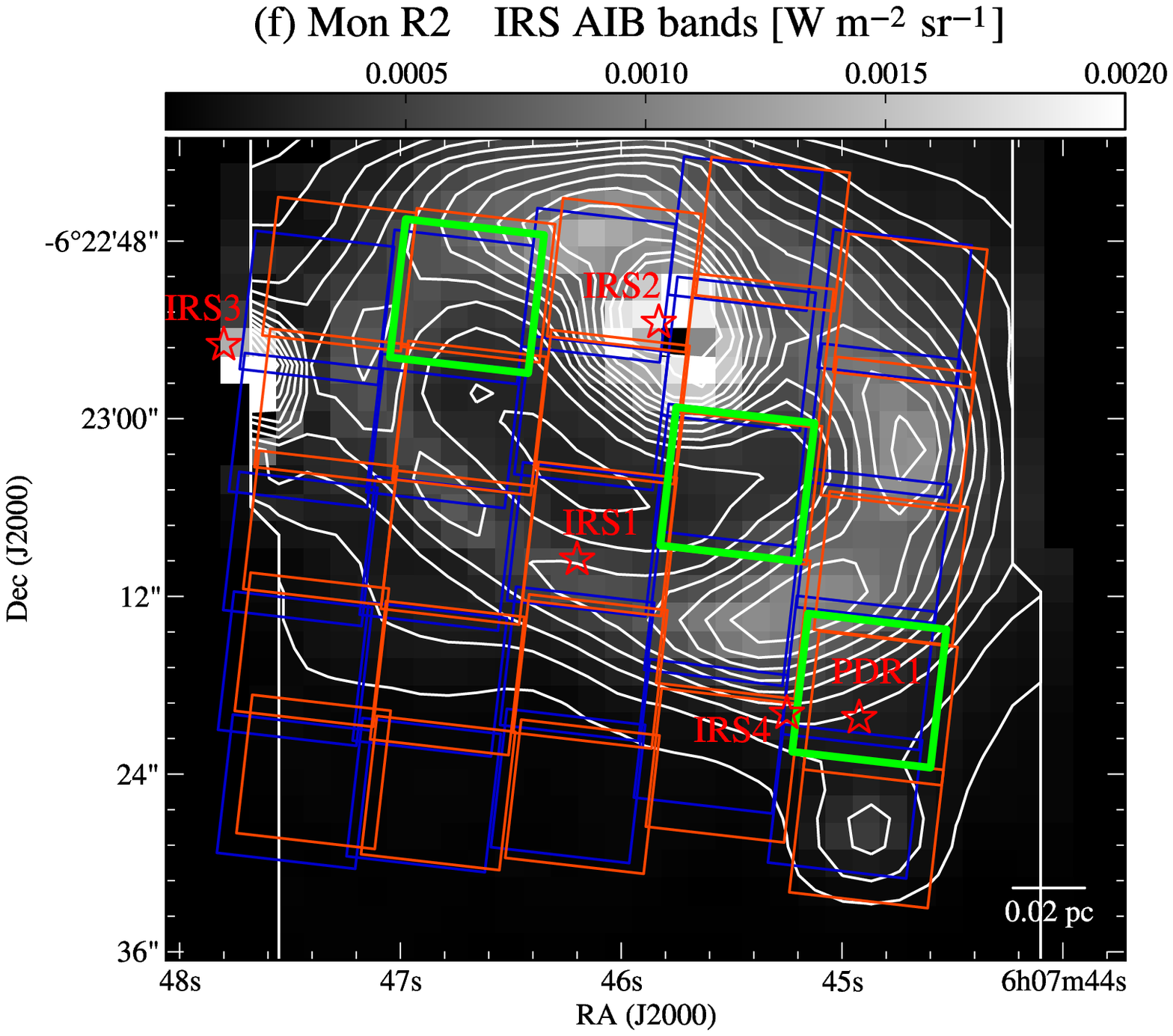}
\caption{PACS footprints of the blue camera (\oi\ 63\um) and the red camera (\oi\ 145\um\ and \cii\ 158\um) overlaid on IRAC 8\um, except for (f) Mon~R2. (There IRAC 8\um\ is saturated, and the total intensity of AIB bands derived from fitting of IRS spectra (see text) is shown.)  Green circles or boxes show the areas from which our PACS and IRS spectra are extracted (subregions in Table~\ref{table:line_intensity}).  For Mon~R2 (f), the positions of four IR sources (IRS1--4) and PDR~1 from \citet{Berne2009} are marked.}
\label{fig:obspos}
\end{figure*}

Table~\ref{table:globalprop} shows the global properties of the investigated regions.  They cover three orders of magnitude in far-ultraviolet (FUV; $h\nu =$ 6--13.6~eV) flux, $G_0$ in units of the Habing field \citep[$1.6\times 10^{-6}$~W\,m$^{-2}$][]{Habing1968}.  Figure~\ref{fig:obspos} shows the PACS field-of-view and the extracted area for each PDR.  Some of these sources have been studied and modeled in detail based on the observations with the Herschel-Heterodyne Instrument for the Far-Infrared \citep[HIFI;][]{degraauw2010} in WADI (see \citeauthor{Joblin2010} \citeyear{Joblin2010} for NGC~7023 and \citeauthor{Fuente2010} \citeyear{Fuente2010} and \citeauthor{Pilleri2012b} \citeyear{Pilleri2012b} for Mon~R2).

\subsubsection{Horsehead}

The Horsehead Nebula emerges from the western edge of L1630 as a dark cloud at visible wavelengths, and its outer edge is delineated by a bright and narrow filament in the MIR emission \citep[Fig.~\ref{fig:obspos}a]{Abergel2003}.  The exciting source $\sigma$ Ori illuminates the filament from the west, and a steep change in column density marks the western edge of the filament, while on the eastern side the MIR emission decreases because of the extinction of the incident radiation by dense material \citep{Abergel2003}.  Using observations of the \hh\ 1-0 S(1) line emission, \citet{Habart2005} proposed a model with a density gradient from $10^4$ to $10^5$~\cc\ with a scale length of about 10\arcsec\ of this filament.  \citet{Goicoechea2009} mapped the \oi\ 63\um\ emission at a low spectral resolution and concluded that a nonlocal and non-LTE treatment can be important to model the \oi\ emission.  They derived a value of \pe\ of 1--2\%.

\subsubsection{Ced~201}

Ced~201 is a reflection nebula, illuminated by the B9.5 star BD $+69^\circ$~1231.  Based on radial velocity measurements, \citet{Witt1987} suggested that this reflection nebula is the result of an accidental encounter of a small dense molecular cloud with an unrelated star.  \citet{Kemper1999} modeled cooling lines of this PDR and concluded that heating by PAHs and VSGs is required to explain the observed cooling-line emissions.  \citet{YoungOwl2002} estimated \pe\ to be $<0.3$\% based on low spatial resolution KAO observations of \cii\ 158\um\ and the upper limit of \oi\ 63\um.

\subsubsection{NGC~7023}

NGC~7023 is a prototype PDR, which has been widely studied at many wavelengths.  The UV radiation from the exciting B star, HD200775, creates three main PDRs in this nebula; the brightest NGC~7023 North-West (hereafter NGC~7023~NW) is located about 40\arcsec\ northwest of the star, another PDR lies about 70\arcsec\ south (NGC~7023~S), and NGC~7023 East (hereafter NGC~7023~E) is located about 170\arcsec\ east of the star \citep{Berne2007}.  \citet{Pilleri2012a} showed a clear difference in the spatial distribution of \pahn\ and \pahp\ in these three PDRs.  \citet{Joblin2010} presented the first results from HIFI observations along a cut through NGC~7023~NW and S, suggesting that both the \cii\ emission and the aromatic infrared-band (AIB) emissions in the MIR arise from the regions located in the transition zone between atomic and molecular gas, providing new insights into the importance of the PAH charge evolution in the energetic studies of PDRs.

\subsubsection{Carina}

The Carina Nebula is a massive star-forming region complex with 65 O-type stars at a distance of 2.3~kpc \citep{Smith2006}, which provides the nearest example of a very massive star-forming region and has been observed at many wavelengths \citep{SmithBrooks2007}.  Trumpler 14 and 16 are the most prominent clusters.  The PDR properties have been investigated using FIR and submilimeter emission lines \citep{Kramer2008,Mizutani2004,Brooks2003}.  The region excited by Trumpler 14 has a FUV flux of $G_0=7\times 10^3$ and a high density of $2\times 10^5$\cc\ \citep{Kramer2008}.  \citet{Preibisch2011} present a large deep 870\um\ continuum map, showing that the total mass is about $2\times 10^5\,M_\odot$.  \citet{Gaczkowski2013} recently published PACS and SPIRE continuum maps of the Carina Nebula complex to study embedded young stellar objects.  In this study, we analyzed a PACS position in Carina North (hereafter Carina~N), excited by Trumpler 14.

\subsubsection{Mon~R2}

Mon~R2 is a close-by ultracompact \hii\ (UC\hii) region at a distance of 850~pc, which comprises several PDRs that can be spatially resolved at both mm and IR wavelengths.  The most intense UV source, which coincides with the infrared source IRS1, is located at the center of the cometary shaped UC\hii.  The brightest PDR of Mon~R2 is illuminated by an extremely intense UV field \citep[$G_0\sim5\times10^5$;][]{Rizzo2003}, and its PAH emission peaks at about 20\arcsec\ northwest of IRS1 (Fig.~\ref{fig:obspos}).  Several other PDRs lie around IRS1 and span different physical conditions such as temperature, density, column density, and UV field \citep[e.g.][]{Rizzo2003,Berne2009,Pilleri2012b}.  All PDRs have very bright MIR spectra, consisting of the emission from AIBs, \hh\ rotational lines, and the continuum.  The spatial extent of these PDRs ($\sim 0.03$~pc) yields an angular size of $\sim 8$\arcsec, as shown by PAH and \hh\ emissions in the MIR \citep{Berne2009}.  The shape of the MIR spectra varies at different PDRs, reflecting the photo-processing of the AIB carriers.


\subsection{Far-infrared spectroscopy with \textit{Herschel}}\label{subsec:reduction_herschel}

\begin{table*}
\caption{PACS observation summary.}
\label{table:pacsobsparam}
\centering
\begin{tabular}{lcrrccl}
\hline
Object && RA (J2000) & DEC (J2000) & Obs. ID & Obs. mode & Lines\\
\hline\hline
Horsehead && $5^\mathrm{h}40^\mathrm{m}53^\mathrm{s}.7$ & $-2^\circ28^\prime00^{\prime\prime}.0$ & 1342228244 & LineSpec, Chop/Nod & \oi\ 63\um\\
&&& & 1342228245 & LineSpec, Chop/Nod & \cii\ 158\um \\
&&& & 1342228246 & LineSpec, Chop/Nod & \oi\ 145\um \\
\hline
Ced~201 && $22^\mathrm{h}13^\mathrm{m}24^\mathrm{s}.6$ & $+70^\circ15^\prime04^{\prime\prime}.6$ & 1342211847 & LineSpec, Chop/Nod & \oi\ 63\um \\
&&&& 1342216208 & LineSpec, Chop/Nod & \oi\ 145\um, \cii\ 158\um\\
\hline
NGC~7023 & E & $21^\mathrm{h}02^\mathrm{m}09^\mathrm{s}.4$ & $+68^\circ09^\prime15^{\prime\prime}.3$ & 1342209729 & LineSpec, Chop/Nod & \oi\ 63\um\\
&&&& 1342222231 & LineSpec, Chop/Nod & \oi\ 145\um, \cii\ 158\um\\
& NW & $21^\mathrm{h}01^\mathrm{m}32^\mathrm{s}.4$ & $+68^\circ10^\prime25^{\prime\prime}.0$ & 1342208910 & LineSpec, Chop/Nod & \oi\ 63\um\\
&&&& 1342222230 & RangeSpec, Chop/Nod & \oi\ 145\um, \cii\ 158\um\\
& NW & $21^\mathrm{h}01^\mathrm{m}35^\mathrm{s}.7$ & $+68^\circ10^\prime01^{\prime\prime}.0$ & 1342196683 & LineSpec, Chop/Nod & \oi\ 63\um\\
&&&& 1342197033 & RangeSpec, Chop/Nod & \oi\ 145\um, \cii\ 158\um\\
\hline
Carina & N & $10^\mathrm{h}43^\mathrm{m}25^\mathrm{s}.7$ & $-59^\circ34^\prime35^{\prime\prime}.3$ & 1342214631 & LineSpec, Unchopped & \oi\ 63\um, \cii\ 158\um\\
\hline
Mon~R2 && $6^\mathrm{h}07^\mathrm{m}46^\mathrm{s}.2$ & $-6^\circ23^\prime08^{\prime\prime}.3$ & 1342228453 & LineSpec, Chop/Nod & \oi\ 63\um\\
&&&& 1342228454/6 & RangeSpec, Unchopped & \oi\ 145\um, \cii\ 158\um\\
\hline
\end{tabular}
\end{table*}

\begin{figure}
\centering
\includegraphics[width=0.45\textwidth]{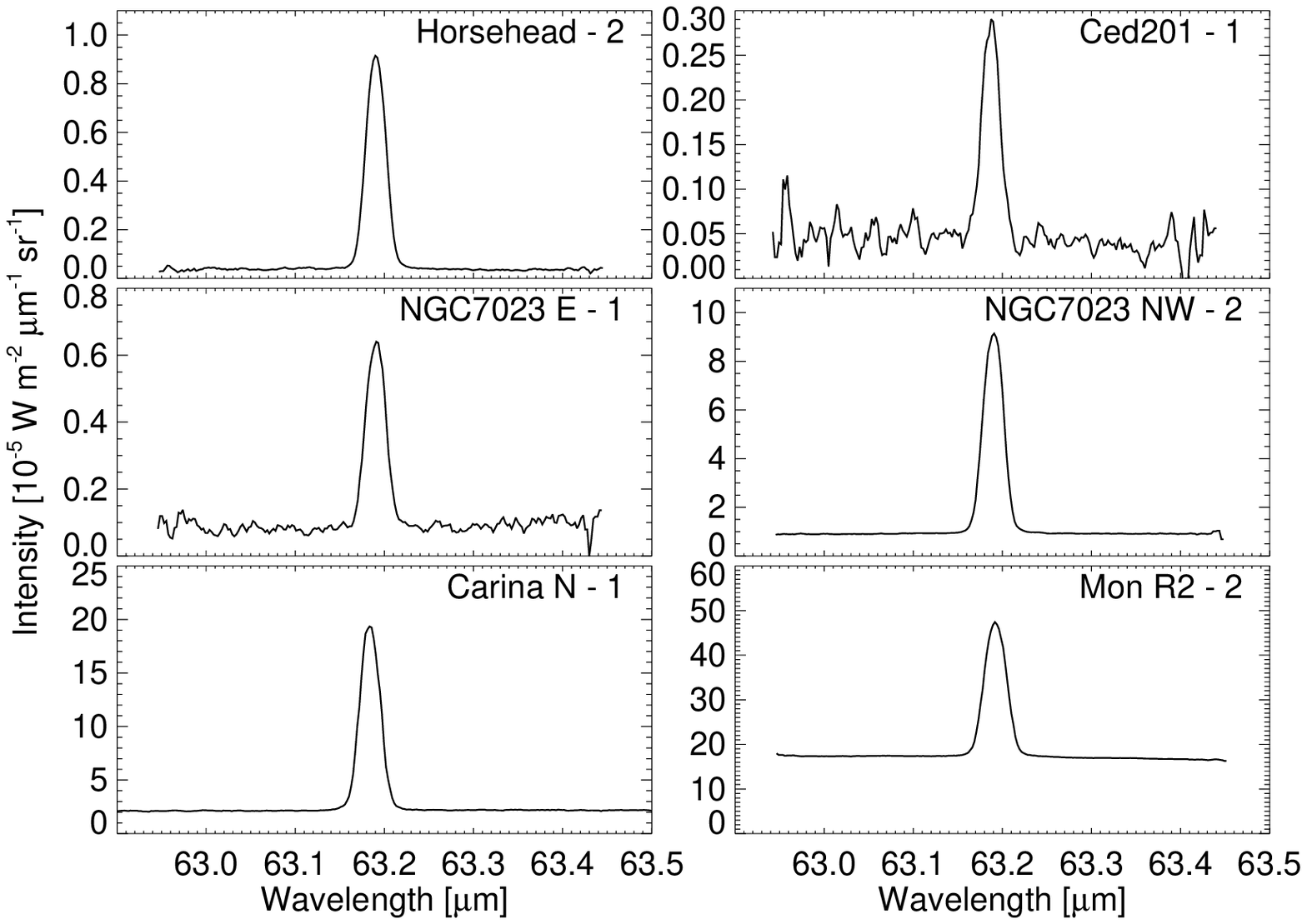}
\includegraphics[width=0.45\textwidth]{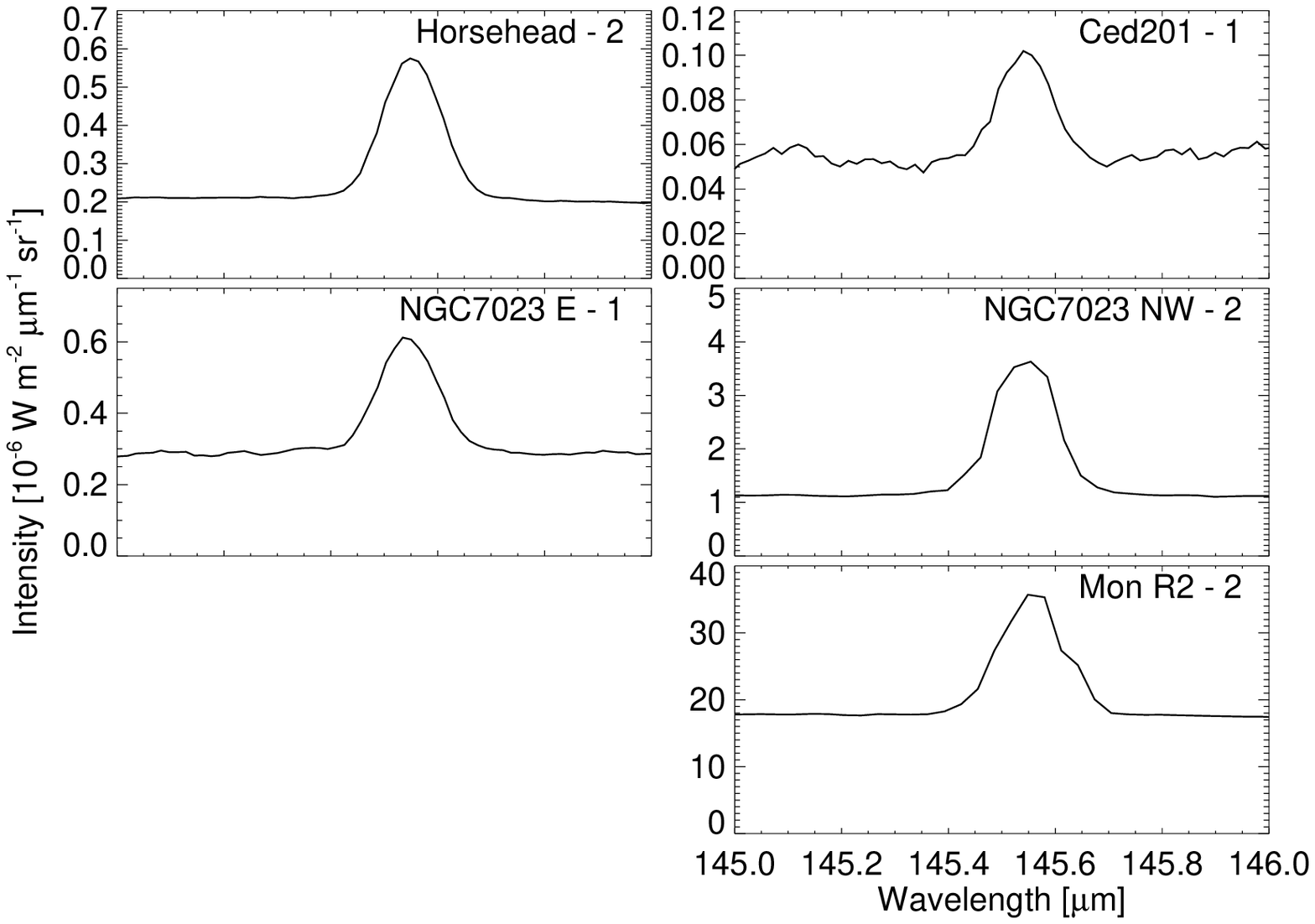}
\includegraphics[width=0.45\textwidth]{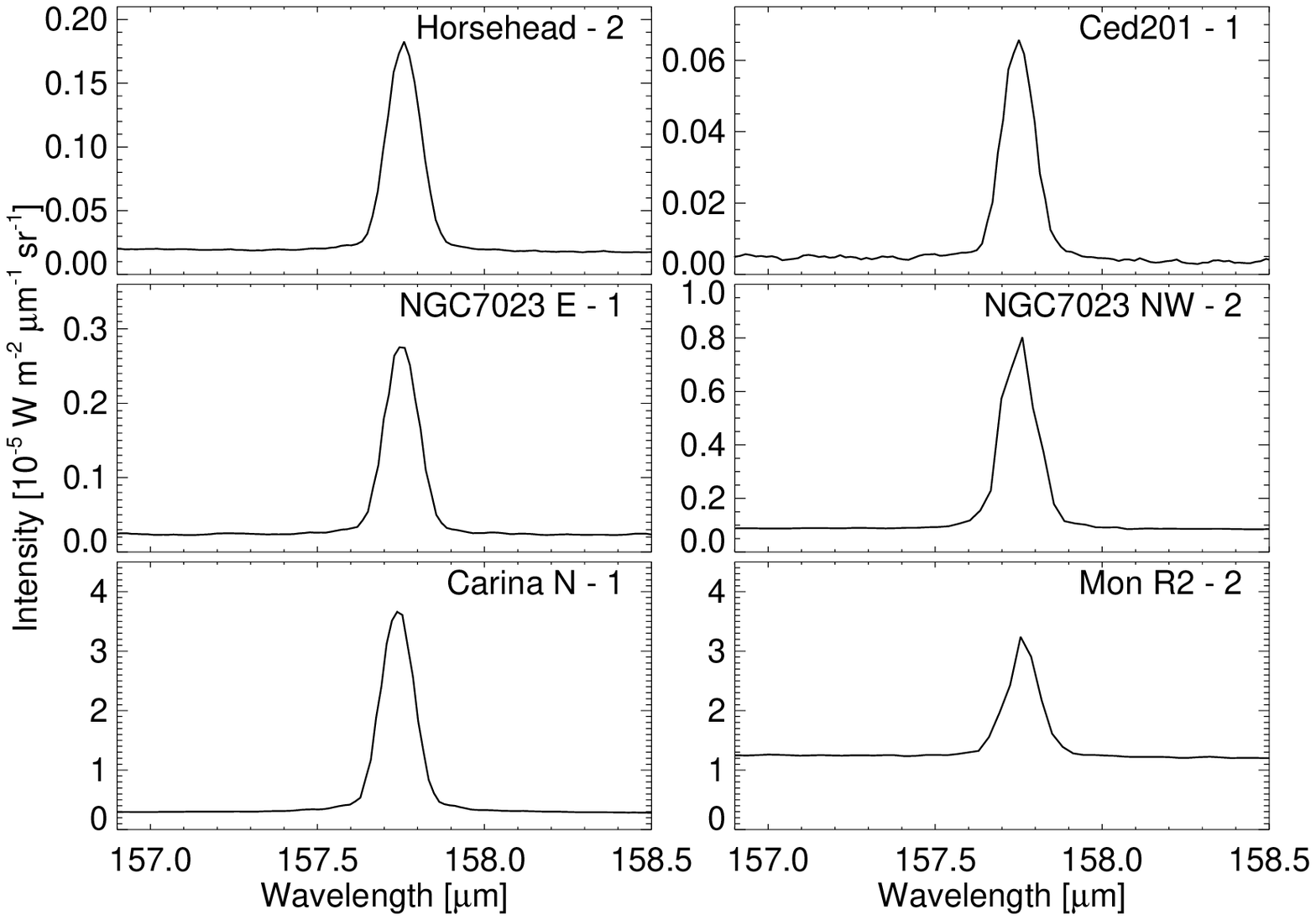}
\caption{Examples of \oi\ 63\um, 145\um, and \cii\ 158\um\ spectra in each PDR.  The numbers labeled in figures after the name of the object show subregions (see Table~\ref{table:line_intensity}).}
\label{fig:line}
\end{figure}

The FIR spectroscopic observations were performed with PACS as part of the WADI program.  The observational parameters are listed in Table~\ref{table:pacsobsparam}.  All measurements are single pointed observations with the PACS spectrometer consisting of a $5\times 5$ array of spatial pixels (spaxels) with a size of $9.4^{\prime\prime}$ each (see Fig.~\ref{fig:obspos}).  We pipelined the data from level 0 to 2 using the Herschel data processing system \citep[HIPE;][]{Ott2010} version 8.  We converted the unit of the pipelined spectra, Jy/spaxel, to W\,m$^{-2}$\,$\mu$m$^{-1}$\,sr$^{-1}$ using the field of view of a spaxel ($9.4^{\prime\prime}\times 9.4^{\prime\prime}$).  For the Chop/Nod observations, we checked the contamination at the OFF positions for \oi\ 63\um, 145\um, and \cii\ 158\um\ using the script provided in HIPE (ChopNodSplitOnOff.py), which provides separate spectra for the ON and OFF positions.  Only the \cii\ 158\um\ emission in the Horsehead shows significant OFF contamination.  Since the northern side of the OFF positions shows stronger \cii\ 158\um, and the emission from the southern side can be attributed to a diffuse component that is not associated with the region, we used only the OFF measurement of the southern side to obtain the flux from the source.  For the unchopped grating-scan observations, some emission is detected at the OFF positions in \cii\ 158\um\ (Carina~N and Mon~R2) and \oi\ 63\um\ (Carina~N).  The line intensities at the OFF positions are 4--7\%, 2--5\%, and 1--4\% of that at the ON positions.  We subtracted the OFF emission in Mon~R2 but not in Carina~N, because the detection of \oi\ 63\um\ in the Carina~N OFF position indicates a real contamination from local dense clouds, while the \cii\ 158\um\ emission traces more diffuse region and can be attributed to the large-scale diffuse Galactic \cii\ emission.  To determine the continuum levels, we subtracted the OFF measurements in all observations to remove the telescope background.

The uncertainty of the absolute flux calibration is $11$--$12$\% for the line and continuum emission\footnote{PACS spectroscopy performance and calibration document v2.4}.  A larger uncertainty is expected in weak sources for unchopped grating scans.  The reproductivity in total absolute flux (telescope + source) of the unchopped grating scan is 4\% (peak-to-peak).  The relative uncertainty against the source flux can be expressed as $0.04\times F(\mathrm{telescope + source}) / F(\mathrm{source})$, where $F$ is the flux.  For Mon~R2, this means an uncertainty of $<13$\% and $<5$\% for the continuum at 105--180\um\ and at the peak of \cii\ 158\um.  For Carina~N, it is $<10$\% and $<5$\% at the peak of the \oi\ 63\um\ and \cii\ 158\um\ emission lines and 20--50\% for the underlying continuum level.

\begin{table*}
\caption{Line intensities observed by PACS.  Subregions are shown by green circles or boxes in Fig.~\ref{fig:obspos}, and are described in detail in Sect.~\ref{subsec:targets}.}
\label{table:line_intensity}
\centering
\begin{tabular}{lccl|ccc}
\hline
Object && \multicolumn{2}{c}{Subregion} & \multicolumn{3}{|c}{Line intensities [W\,m$^{-2}$\,sr$^{-1}$]} \\
&&&& \oi\ 63\um & \oi\ 145\um & \cii\ 158\um \\
\hline
Horsehead &  & 1 & \hii\ region side & $(1.01 \pm 0.02)\times 10^{-7}$ & $(1.28 \pm 0.03)\times 10^{-8}$ & $(9.49 \pm 0.23)\times 10^{-8}$ \\
& & 2 & along the ridge & $(2.39 \pm 0.02)\times 10^{-7}$ & $(5.08 \pm 0.03)\times 10^{-8}$ & $(2.27 \pm 0.02)\times 10^{-7}$ \\
& & 3 & molecular side & $(1.29 \pm 0.02)\times 10^{-7}$ & $(2.75 \pm 0.03)\times 10^{-8}$ & $(1.66 \pm 0.02)\times 10^{-7}$ \\
\hline
Ced~201 &  & 1 & 9.4\arcsec\ diameter & $(6.02 \pm 0.24)\times 10^{-8}$ & $(5.48 \pm 0.47)\times 10^{-9}$ & $(7.55 \pm 0.09)\times 10^{-8}$ \\
& & 2 & 28.2\arcsec\ diameter & $(4.59 \pm 0.13)\times 10^{-8}$ & $(4.33 \pm 0.24)\times 10^{-9}$ & $(6.09 \pm 0.04)\times 10^{-8}$ \\
\hline
NGC~7023 & E  & 1 & 9.4\arcsec\ diameter & $(1.53 \pm 0.03)\times 10^{-7}$ & $(4.48 \pm 0.07)\times 10^{-8}$ & $(3.23 \pm 0.03)\times 10^{-7}$ \\
& & 2 & 28.2\arcsec\ diameter & $(1.56 \pm 0.02)\times 10^{-7}$ & $(3.90 \pm 0.03)\times 10^{-8}$ & $(3.04 \pm 0.02)\times 10^{-7}$ \\
\hline
NGC~7023 & NW  & 1 & cavity & $(6.90 \pm 0.06)\times 10^{-7}$ & $(6.66 \pm 0.21)\times 10^{-8}$ & $(5.53 \pm 0.11)\times 10^{-7}$ \\
& & 2 & interface & $(2.31 \pm 0.02)\times 10^{-6}$ & $(3.73 \pm 0.09)\times 10^{-7}$ & $(9.09 \pm 0.24)\times 10^{-7}$ \\
& & 3 & molecular region & $(1.47 \pm 0.01)\times 10^{-6}$ & $(3.05 \pm 0.05)\times 10^{-7}$ & $(6.59 \pm 0.10)\times 10^{-7}$ \\
\hline
Carina & N  & 1 & $8$\um\ clump & $(4.89 \pm 0.04)\times 10^{-6}$ &  --  & $(4.46 \pm 0.04)\times 10^{-6}$ \\
& & 2 & east-west ridge & $(4.00 \pm 0.02)\times 10^{-6}$ &  --  & $(4.10 \pm 0.03)\times 10^{-6}$ \\
& & 3 & southern $5\times 3$ spaxels & $(3.51 \pm 0.02)\times 10^{-6}$ &  --  & $(3.98 \pm 0.03)\times 10^{-6}$ \\
\hline
Mon~R2 &  & 1 & around PDR1 & $(7.68 \pm 0.08)\times 10^{-6}$ & $(1.66 \pm 0.06)\times 10^{-6}$ & $(1.64 \pm 0.07)\times 10^{-6}$ \\
& & 2 & southwest & $(9.37 \pm 0.12)\times 10^{-6}$ & $(3.56 \pm 0.14)\times 10^{-6}$ & $(3.06 \pm 0.12)\times 10^{-6}$ \\
& & 3 & northeast & $(9.05 \pm 0.17)\times 10^{-6}$ & $(2.38 \pm 0.09)\times 10^{-6}$ & $(2.73 \pm 0.10)\times 10^{-6}$ \\
\hline
\end{tabular}
\end{table*}

We defined the region to be used in our study depending on the morphology of each source so that the widest range of physical conditions is included.  Since our PACS observations are not fully sampled, we cannot exploit the entire spatial information.  Instead, we selected a few typical areas in each source and extracted the \oi\ and \cii\ line intensities and the continuum flux from PACS observations and combined them with the IRS results.  These areas are listed in Table~\ref{table:line_intensity}, shown as green boxes or circles in Fig.~\ref{fig:obspos}, and are explained in detail in the following.  We extracted spectra based on the geometrical area and did not apply a beam-size correction.  The FWHM of the PACS spectrometer beam at 63\um\ and 158\um\ is $\sim 9^{\prime\prime}$ and $\sim 11.5^{\prime\prime}$, respectively, and the uncertainty from the difference of the beam size is estimated to be $\sim 15$\% (see Appendix~\ref{sect:app_beamsize}).

For Horsehead, Carina~N, and Mon~R2, all observations were made with almost the same position angle, i.e., the PACS spaxels observed almost the same area in the sky for different wavelengths (Fig.~\ref{fig:obspos}).  In this case, we extracted the areas to be examined on the basis of the PACS spaxels.  In the Horsehead, the spaxels are aligned to the ridge.  We defined three areas; the first one lies at the western side of the ridge in the ionized gas (an average over four spaxels except for the north-end spaxel), the second area is along the ridge (over five spaxels), and the third area is in molecular gas, at the eastern side of the ridge (five spaxels).  In Carina~N, we selected three regions; a single spaxel toward a clump seen in the IRAC 8\um\ map, a ridge-like east-west structure in the 8\um\ map including this clump (five spaxels), and the southern $5\times 3$ spaxels where the PAH emissions in the IRS spectra are prominent.  In Mon~R2, we selected three spaxels around PDR1 \citep[following the definition in][]{Berne2009}, where the \pahp\ fraction is suggested to be low, and toward the southwest and northeast inner edges of the \pahp\ distributions shown in \citet{Berne2009}.  They correspond to the three green boxes in Fig.~\ref{fig:obspos}f from southwest to northeast.

For Ced~201, NGC~7023~E, and NGC~7023~NW, the \oi\ 63\um\ line was observed in different seasons of the year than the \oi\ 145\um\ and \cii\ 158\um\ observations, and the difference of the position angle is significant.  Therefore, we defined circles as areas to be studied, and took a weighted mean of PACS spectra based on the geometrical overlap between the defined circles and the PACS spaxels.  For Ced~201 and NGC~7023~E, two circles were defined with a common center for both PACS footprints and diameters of $9.4$\arcsec\ and $28.2$\arcsec.  For NGC~7023~NW, we selected three typical regions that trace different ionization fractions of PAHs based on the analysis in \citet{Pilleri2012a}.  The areas were defined by circles with 12\arcsec\ diameter, at the interface with strong PAH emission, toward the molecular cloud, and in the cavity toward the exciting star (Fig.~\ref{fig:obspos}).

After extracting the spectra in each region as described above, we obtained the \oi\ 63\um\ and \cii\ 158\um\ intensities, as well as \oi\ 145\um\ when available, by a Gaussian fit after linear baseline subtraction (Fig.~\ref{fig:line}, Table~\ref{table:line_intensity}).  In general, the ratio of \oi\ 63\um\ / \cii\ 158\um\ traces the density of the PDRs \citep{Roellig2006}.  Among our PDRs, NGC~7023~E and Ced~201 show a low ratio ($<1$), indicating a low density, and Mon~R2 has the highest ratio ($\gtrsim 4$).  This trend is roughly consistent with the density estimate in previous studies (Table~\ref{table:tir_and_uv}, Appendix~\ref{sect:app_gamma}).  The \oi\ 145\um/63\um\ is $>0.09$ for all targets and $0.3$ at maximum, which exceeds the PDR model prediction for the relevant physical parameter ranges \citep{Roellig2006,Kaufman1999}.  Possible reasons are high optical depths of the \oi\ 63\um\ emission, a suprathermal population of \oi\ by collision with \hh, and foreground absorption in \oi\ 63\um\ \citep{Liseau2006}.  Although the optical depth of the \oi\ 63\um\ line is taken into account by the PDR models for simple geometries, the overlapped several PDR clumps along the line-of-sight or the edge-on geometry can cause more significant self-absorption of \oi\ 63\um\ \citep{Habart2003,Okada2003}, and the non-local calculation suggests a higher \oi\ 145\um/63\um\ ratio \citep{Elitzur2006}.  The foreground absorption cannot be quantified in our PDRs because of the lack of velocity-resolved \oi\ 63\um\ observations.  However, the HIFI observations of the velocity-resolved \cii\ 158\um\ imply only minor effects toward our PDRs.

\subsection{Mid-infrared spectroscopy with \textit{Spitzer}}\label{subsec:obs_sst}

We analyzed the MIR spectra observed with the Short-Low (SL1 and SL2) module of IRS onboard \textit{Spitzer} except for NGC~7023~E, for which we used the ISOCAM highly-processed data product \citep{Boulanger2005}.  We used the same data cubes as in \citet{Pilleri2012a} for the Horsehead, Ced~201, and NGC~7023, and the data reduction is described in that paper as well.  For Mon~R2, the IRS data and their reduction are shown in \citet{Berne2009}.  For Carina~N, we collected the IRS observations from the \textit{Spitzer} data archive and analyzed them using CUBISM \citep{Smith2007}.  To estimate the continuum, Short-High (SH) and Long-Low (LL) spectra were also reduced and extracted when available.  Their intensities are scaled to match the SL spectra, then the whole spectra are scaled to match the photometric data; IRAC 8\um\ and MIPS 24\um\ for Horsehead, Ced~201, and NGC7023~E, and IRAC 8\um\ for NGC7023~NW and Carina~N.  For Mon~R2, the absolute flux scaling was not applied because IRAC 8\um\ is saturated, and the available spectral range does not cover the MIPS 24\um\ band.  The correction should be less significant in Mon~R2 because it is the brightest source in our samples.  Applying this correction does not affect any trend in the following results and changes none of our conclusions.  We extracted a spectrum from the same area as for the PACS spectra described in the previous subsection.

\section{Analysis}
\subsection{Total infrared intensity}\label{subsec:analysis_fir}

\begin{figure*}
\centering
\includegraphics[width=0.9\textwidth]{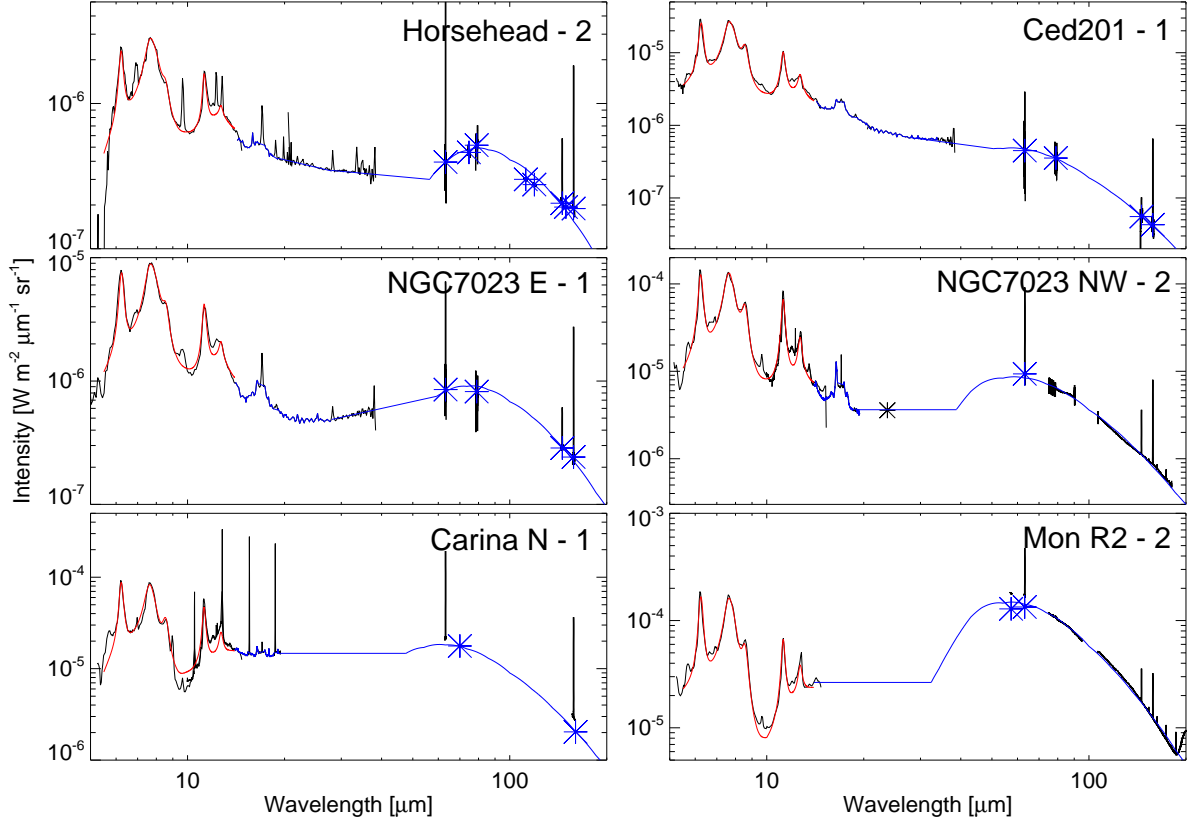}
\caption{Continuum fit at the same positions as in Fig.~\ref{fig:line}.  The black lines are the observed spectra, while the red and blue lines are the models used to estimate the TIR.  The blue asterisks are the data points that are used for fitting the FIR thermal dust emission (see text).  The blue curve in the FIR is the fit with the thermal dust model, that in the MIR indicates either the direct integration, the linear fit, or the assumed constant value.  The red curve below 14\um\ shows the fit described in Sect.~\ref{subsec:analysis_pah}.  The black asterisk in NGC7023~NW is the MIPS 24\um\ flux.}
\label{fig:tir}
\end{figure*}

Observationally, \pe\ is measured as (\oi\ $+$ \cii) / TIR, where TIR is the total infrared flux \citep[3--1100\um; ][]{DaleHelou2002}, representing the fraction of the input energy that is converted into the cooling lines.  The total FIR flux, 42.5--122.5\um\ \citep{Helou1988} or the integrated flux of the thermal emission by large dust grains, is often used as a tracer of the input energy.  Here we used the TIR instead, which is the emission from all dust grains, including PAHs and VSGs, to represent the total input energy.  As dominant cooling lines, we summed the intensities of \oi\ 63\um, 145\um, and \cii\ 158\um, except for Carina~N, where the \oi\ 145\um\ was not observed, and we used (\oi\ 63\um\ $+$ \cii\ 158\um).  A second PACS observation 1.3$^\prime$ away in Carina~N, which is not used in this study because it does not overlap with the IRS observations, shows an \oi\ 145\um/63\um\ intensity ratio of $0.05$--$0.15$.  If our position has the same ratio, neglecting the \oi\ 145\um\ emission underestimates \pe\ by $8$\% at most.

To estimate TIR as the energy input tracer we need to integrate the grain emission of 3--1100\um.  The contribution from 3--5.5\um\ is negligible because of the low intensity at 5.5\um\ in all regions.  To obtain the energy in the range of 5.5--14\um, we fit the IRS spectra using the procedure described in \citet{Pilleri2012a} and integrated the resulting fit (see Sect.~\ref{subsec:analysis_pah}).

For $\lambda > 14$\um, the available data and the quality are not uniform, and we fine-tuned approaches from region to region to estimate the integrated flux (see Fig.~\ref{fig:tir}), although the basic approach was the same; we fit the FIR spectra from PACS observations with the thermal dust model, extrapolated the fit in the MIR to longer wavelengths, and connected the two fits at the intersection point.

In Horsehead, Ced~201, and NGC~7023~E, MIR spectra are available up to 35\um.  We estimated the integrated flux for 14--28\um\ by direct integration excluding the strong emission lines for Ced~201 and NGC~7023~E, and by a linear fit to the spectra in the Horsehead because of the lower signal-to-noise ratio (S/N).  The 28--32\um\ range is fitted by a linear function, and the results are extrapolated to longer wavelengths.  In NGC~7023~NW and Carina~N, IRS/SH data are available for 14--19\um, which are directly integrated, and we assumed a flat spectrum from 19\um\ up to the wavelength where the big-grain emission, peaking in the FIR, exceeds this level.  For Mon~R2, we assumed that the flux at $\lambda > 14$\um\ is constant (Fig.~\ref{fig:tir}).

For the PACS Chop/Nod LineSpec observations (Table~\ref{table:pacsobsparam}), the continuum levels are available around the emission lines.  They are indicated as blue asterisks in Fig.~\ref{fig:tir}.  In NGC~7023~NW and Mon~R2, the full spectral data from the PACS RangeSpec observations at $>70$\um, after excluding strong emission lines, were used to fit the FIR continuum.  In Carina~N, the uncertainty of the absolute continuum flux is too large (see Sect.~\ref{subsec:reduction_herschel}).  Therefore we obtained the photometry data at blue (70\um) and red (160\um) bands from the \textit{Herschel} data archive.  We fit the FIR continuum using the dust emissivity of \citet{Ossenkopf1994}, for a gas density of $10^6$~\cc\ and grains with thin ice mantles, considering the temperature and the column density as free parameters.  Since we discuss only the integrated IR flux, the choice of the dust emissivity is not critical.  Where this fit was exceeded by the extrapolation from the MIR spectrum, we connected these two at the intersection point.  The final fits are shown as blue lines in Fig.~\ref{fig:tir}.

Since the PACS observations are not fully sampled and the IRS spectra were extracted just using the geometrical area, the difference of the PSF could affect the resulting fit of the spectral energy distributions (SEDs).  However, the uncertainty decreases for larger regions (average over several spaxels) and it does not change the general trend of \pe\ (Sect.~\ref{sec:discussion}; Fig.~\ref{fig:correlation}).

In NGC~7023~NW, the MIPS 24\um\ flux is available but was not used to scale the IRS spectra because the spectral range of the IRS does not cover the MIPS 24\um\ band.  Figure~\ref{fig:tir} shows that the MIPS 24\um\ flux matches the assumed constant continuum flux.

We translated the obtained TIR into the impinging FUV flux.  By assuming that the entire FUV energy is absorbed by dust grains and reradiated as infrared flux, we estimate $G_0$(TIR) as $4\pi \times \mathrm{TIR}/(1.6\times 10^{-6})$ (Table~\ref{table:tir_and_uv}).  The systematic uncertainty of TIR is $\sim 15$\%, which comes from difference of the beam sizes at different wavelengths (Sect.~\ref{subsec:reduction_herschel}).  In this formula we assumed a spherical geometry.  When we resolve an edge-on PDR with a shorter thickness compared to the length of the line-of-sight, $G_0$(TIR) is overestimated \citep{Meixner1992}.  Another uncertainty of $G_0$(TIR) is the contribution of photons with an energy outside of 6~eV $< h\nu <$ 13.6~eV to the thermal heating of the dust.  For later B-type stars, the contribution of $<6$~eV becomes significant, which causes the overestimate of $G_0$(TIR).  The SED of a B9 star by \citet{Castelli2004} indicates that only 1/4--1/5 of the total energy is emitted in the range of 6~eV $< h\nu <$ 13.6~eV from the star.  We did not apply the correction and treated $G_0$(TIR) as an upper limit.  Detailed comparisons of $G_0$(TIR) with previous studies and other diagnostics are described in Appendix~\ref{sect:app_gamma}, and the final $G_0$ estimates, which are used in the following analysis, are also listed in Table~\ref{table:tir_and_uv}.

\begin{table*}
\caption{TIR and $G_0$(TIR) and the parameters used to estimate the charging parameter $\gamma=G_0 T^{1/2}/n_e$.  The uncertainty of TIR and $G_0$(TIR) is $\sim 15$\% (see text).}
\label{table:tir_and_uv}
\centering
\begin{tabular}{lcl|c|c|ccc}
\hline
Object && Sub & TIR & $G_0$(TIR) & \multicolumn{3}{|c}{Adopted parameters for $\gamma$}\\
&&region& [W\,m$^{-2}$\,sr$^{-1}$] & [Habing field] & $G_0$ & \nh\ [\cc] & T [K] \\
\hline
Horsehead & &1 & $2.7\times 10^{-5}$ & $210$ & $100$--$210$ & ($0.3$--$2$)$\times 10^5$ & $234$--$550$\\
&&2 & $7.0\times 10^{-5}$ & $550$ & $100$--$550$ & ($0.1$--$2$)$\times 10^5$ & $234$--$550$\\
&&3 & $6.0\times 10^{-5}$ & $470$ & $100$--$470$ & ($0.1$--$2$)$\times 10^5$ & $234$--$550$\\
\hline
Ced~201 & &1 & $1.2\times 10^{-4}$ & $940$ & $200$--$940$ & $4\times 10^2$--$1.2\times 10^4$ & $330$\\
&&2 & $7.6\times 10^{-5}$ & $600$ & $200$--$600$ & $4\times 10^2$--$1.2\times 10^4$ & $330$\\
\hline
NGC~7023 & E &1 & $1.2\times 10^{-4}$ & $970$ & $120$--$970$ & ($0.1$--$1$)$\times 10^4$ & $258$--$370$\\
&&2 & $1.1\times 10^{-4}$ & $900$ & $120$--$900$ & ($0.1$--$1$)$\times 10^4$ & $258$--$370$\\
\hline
NGC~7023 & NW &1 & $6.3\times 10^{-4}$ & $5.0\times 10^{3}$ & ($0.5$--$1$)$\times 10^4$ & $150\pm 100$ & $430$--$450$\\
&&2 & $9.8\times 10^{-4}$ & $7.7\times 10^{3}$ & ($2.6$--$7.7$)$\times 10^3$ & ($0.01$--$2$)$\times 10^5$ & $430$--$450$\\
&&3 & $6.9\times 10^{-4}$ & $5.4\times 10^{3}$ & ($0.5$--$5.4$)$\times 10^3$ & ($0.2$--$2$)$\times 10^5$ & $430$--$450$\\
\hline
Carina & N &1 & $2.0\times 10^{-3}$ & $1.6\times 10^{4}$ & ($0.7$--$1.6$)$\times 10^4$ & ($0.2$--$1$)$\times 10^6$ & $296$--$529$\\
&&2 & $2.0\times 10^{-3}$ & $1.5\times 10^{4}$ & ($0.8$--$1.5$)$\times 10^4$ & ($0.2$--$1$)$\times 10^6$ & $292$--$440$\\
&&3 & $1.9\times 10^{-3}$ & $1.5\times 10^{4}$ & ($0.8$--$1.5$)$\times 10^4$ & ($0.2$--$1$)$\times 10^6$ & $290$--$390$\\
\hline
Mon~R2 & &1 & $1.0\times 10^{-2}$ & $8.2\times 10^{4}$ & ($5.2$--$8.2$)$\times 10^4$ & ($0.4$--$4$)$\times 10^5$ & $314$--$574$\\
&&2 & $1.1\times 10^{-2}$ & $8.5\times 10^{4}$ & ($0.9$--$1.0$)$\times 10^5$ & ($0.4$--$4$)$\times 10^5$ & $314$--$574$\\
&&3 & $1.3\times 10^{-2}$ & $1.0\times 10^{5}$ & ($1.0$--$1.2$)$\times 10^5$ & ($0.4$--$4$)$\times 10^5$ & $314$--$574$\\
\hline
\end{tabular}
\end{table*}

\subsection{Ionization of PAHs}\label{subsec:analysis_pah}

\begin{figure*}
\centering
\includegraphics[width=0.9\textwidth]{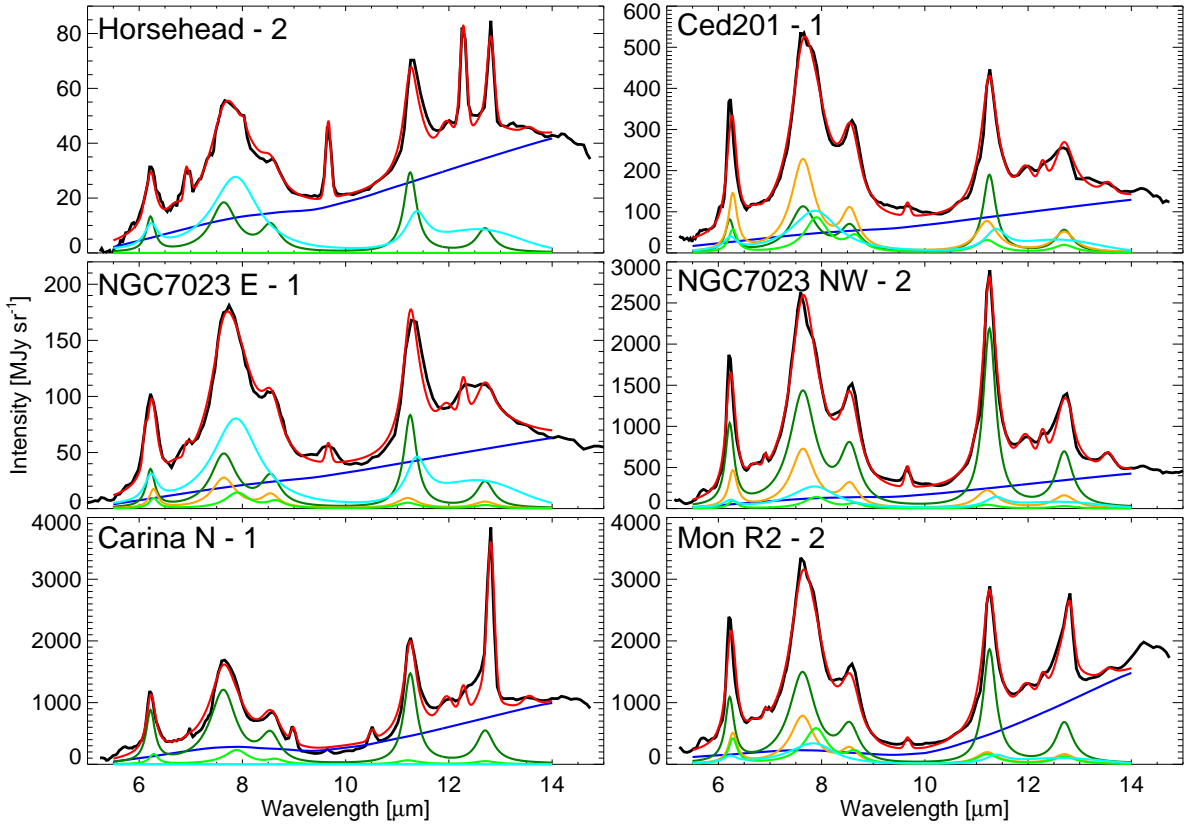}
\caption{Fit of the MIR spectra using the PAHTAT procedure \citep{Pilleri2012a} at the same positions as in Fig.~\ref{fig:line}, assuming the mixed extinction model and $R_V=3.1$.  The black line shows the observed spectrum, the red line is the fitted spectrum, the blue line represents the continuum, green, orange, light-green and light-blue show the \pahn, \pahp, \pahx, and eVSG components.}
\label{fig:irsfit}
\end{figure*}

\begin{table*}
\caption{The fraction of each component relative to the total band emission in the PAH fit.  The variation in each column describes the range covered by the different assumptions in the fit.}
\label{table:pah_fitting}
\centering
\begin{tabular}{lcc|cccc}
\hline
Object && sub & \multicolumn{4}{c}{Fraction of each component} \\
&& region & \pahn & \pahp & \pahx & eVSG\\
\hline
Horsehead &  & 1 & $1.00$ & $0.00$ & $0.00$ & $0.00$ \\
& & 2 & $0.39$ & $0.00$ & $0.00$ & $0.61$ \\
& & 3 & $0.42$ & $0.00$ & $0.00$ & $0.58$ \\
Ced~201 &  & 1 & $0.21$--$0.25$ & $0.38$--$0.44$ & $0.00$--$0.14$ & $0.23$--$0.34$ \\
& & 2 & $0.19$--$0.22$ & $0.21$--$0.25$ & $0.00$--$0.10$ & $0.48$--$0.56$ \\
NGC~7023 & E  & 1 & $0.28$--$0.30$ & $0.13$--$0.15$ & $0.00$--$0.07$ & $0.51$--$0.57$ \\
& & 2 & $0.28$--$0.30$ & $0.13$--$0.15$ & $0.00$--$0.08$ & $0.49$--$0.56$ \\
NGC~7023 & NW  & 1 & $0.13$ & $0.86$ & $0.00$ & $0.00$ \\
& & 2 & $0.59$--$0.61$ & $0.23$--$0.27$ & $0.00$--$0.05$ & $0.11$--$0.15$ \\
& & 3 & $0.59$--$0.60$ & $0.03$ & $0.00$--$0.01$ & $0.37$--$0.38$ \\
Carina & N  & 1 & $0.87$--$0.94$ & $0.00$ & $0.00$--$0.13$ & $0.00$--$0.11$ \\
& & 2 & $0.88$--$1.00$ & $0.00$ & $0.00$--$0.12$ & $0.00$--$0.03$ \\
& & 3 & $0.88$--$1.00$ & $0.00$ & $0.00$--$0.12$ & $0.00$--$0.02$ \\
Mon~R2 &  & 1 & $0.41$--$0.54$ & $0.19$--$0.35$ & $0.00$--$0.18$ & $0.07$--$0.25$ \\
& & 2 & $0.14$--$0.25$ & $0.53$--$0.67$ & $0.00$--$0.20$ & $0.02$--$0.20$ \\
& & 3 & $0.12$--$0.24$ & $0.51$--$0.67$ & $0.00$--$0.22$ & $0.03$--$0.21$ \\
\hline
\end{tabular}
\end{table*}

To derive the contribution of \pahp\ in each region, we applied the PAHTAT (PAH Toulouse Astronomical Templates) procedure described in \citet{Pilleri2012a}.  This procedure fits MIR spectra using a minimal set of template spectra.  The PAH-related templates comprise \pahn, \pahp, larger ionized PAHs \citep[named \pahx, see][]{Joblin2008}, and evaporating very small grains (eVSGs).  In the MIR the \pahn\ and \pahp\ templates are characterized by a strong difference in the relative strength of their 7.7\um\ and 11.3\um\ band features; \pahp\ has a stronger 7.7\um\ band.  The \pahx\ population consists of large ionized PAHs, which was introduced to provide a better fit to planetary nebula spectra: its MIR spectrum is similar to \pahp, but the 7.7\um\ band is shifted to longer wavelengths \citep{Joblin2008,Pilleri2012a}.  eVSGs are an intermediate population between PAHs and classical VSGs.  They present both a broad band and continuum emission in the MIR range.  The PAHTAT procedure also allows simultaneous fitting of the gaseous emission lines, underlying continuum, and the extinction by dust grains along the line-of-sight.  More details of the fitting tool can be found in \citet{Pilleri2012a}.  We assumed a linear continuum except for Mon~R2, where a continuum of two slopes connected at 10\um\ was adopted since big grains contribute significantly to the MIR continuum because of the strong UV field.  To explore the parameter space of the fitting, we used all combinations of the following different assumptions: the dust extinction curves with $R_V=3.1$ and $5.5$, geometries in which the absorbing materials are placed in the foreground or are fully mixed with the emitting materials, and since the presence of \pahx\ in PDRs with an intermediate UV radiation field strength is doubtful, we also tested the results including or excluding the \pahx\ component from the fit.  The difference between these different assumptions is included into the uncertainty of the results, although they are typically smaller than the errors estimated by the discrepancy between the observed spectra and the model (see below).  The fitted spectra are shown in Fig.~\ref{fig:irsfit}, and the derived fraction of each PAH and eVSG contribution to the total band flux is shown in Table~\ref{table:pah_fitting}.

In the following we define the fraction of positively ionized PAHs by the integrated intensity ratio as $f(+)=\int I_{\mathrm{PAH}^+}/(\int I_{\mathrm{PAH}^0}+\int I_{\mathrm{PAH}^+})$, where the integration is made over the wavelength from $5.5$\um\ to $14$\um.  The uncertainty of this fraction can be attributed to two different effects: (1) fitting errors and (2) systematic errors in the assumptions within PAHTAT, i.e., due to intrinsic uncertainties in the templates.  To estimate the first contribution, we integrated the absolute discrepancy of the observed spectra from the model.  We assumed that this residual consists of the uncertainty of the band emissions of small grains (\pahn, \pahp, \pahx, and eVSGs) and the continuum emissions; $\int\mathrm{(residual)}=\Delta\mathrm{(band)}+\Delta\mathrm{(continuum)}$, and the ratio of the uncertainty $\Delta\mathrm{(band)}/\Delta\mathrm{(continuum)}$ is the same as the ratio of the integrated flux $\int I_\mathrm{band}/\int I_\mathrm{continuum}$.  Thus, the uncertainty of the integrated band emission is defined as $\Delta\mathrm{(band)}=\int\mathrm{(residual)}\times \int I_\mathrm{band}/(\int I_\mathrm{band}+\int I_\mathrm{continuum})$.  Then we attribute it to the uncertainty of the \pahp\ flux, i.e., the final uncertainty of $f(+)$ is expressed as $\Delta\mathrm{(band)}/(\int I_{\mathrm{PAH}^0}+\int I_{\mathrm{PAH}^+})$.  This is conservative, because it assumes that the entire uncertainty in the band emission comes from that of \pahp.

The second uncertainty in the obtained $f(+)$ comes from the templates for \pahn\ and \pahp.  The templates are constructed based on observational data and a mathematical blind signal separation (BSS) method \citep{Pilleri2012a}.  The spectral properties of the BSS-extracted spectra, templates, and their assignment to \pahp\ and \pahn\ have been discussed in \citet{Rapacioli2005}, \citet{Berne2007}, \citet{Joblin2008}, and \citet{Berne2009}.  Recently, \citet{Rosenberg2011} compared the BSS extraction to the theoretical spectra with the density functional theory and showed good agreement, although this is for a spectral range of 10--19.5\um.  A precise determination of the uncertainty is difficult, however, given the nature of the extraction method and our limited knowledge of the PAH populations that exist in space.  Using a Monte Carlo approach, \citet{Rosenberg2011} showed in their BSS extraction a $1\sigma$ uncertainty of typically 10\%\ and up to 30\%\ in some parts of the spectrum (see their Fig.~3).  However, this uncertainty in the templates systematically propagates to the uncertainty of $f(+)$ and does not change the trend and our conclusion discussed below.  Therefore, we present our errorbars in the following figures based on the fitting error described above.

\section{Results and discussion}\label{sec:discussion}

\subsection{Relation between \pe\ and $f(+)$}

\begin{figure}
\centering
\includegraphics[width=0.49\textwidth]{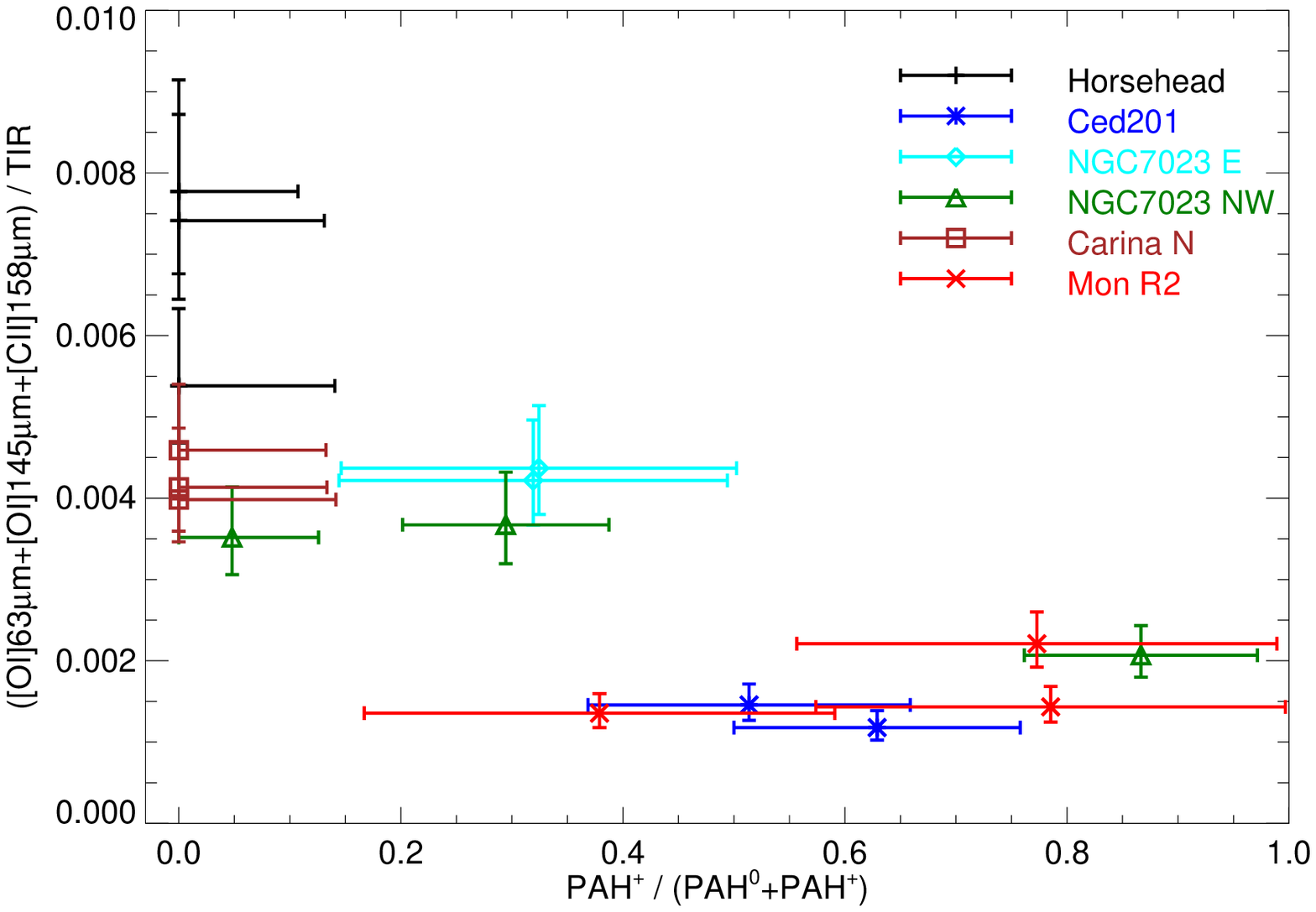}
\caption{\pe $=$ (\oi\ 63\um\ $+$ \oi\ 145\um\ $+$ \cii\ 158\um) / TIR versus the fraction of ionized PAHs ($f(+)$).}
\label{fig:correlation}
\end{figure}

Figure~\ref{fig:correlation} shows the relation between the \pe\ = (\oi\ 63\um\ $+$ \oi\ 145\um\ $+$ \cii\ 158\um) / TIR and the fraction of positively ionized PAHs $f(+)$.  Both \pe\ and $f(+)$ strongly vary among the regions; \pe\ varies from 0.1\% to 0,9\%, and $f(+)$ from $0$ ($+11$)\% to $87$ ($\pm 10$)\%.  All positions with a high $f(+)$ show a low \pe, and all positions with a high \pe\ show a low $f(+)$.  This trend supports the theoretical expectation in which \pe\ decreases when grains are positively ionized, because the energy required to eject electrons from positively charged grains is higher than that from neutral grains.  Here we take the ionization of PAHs as a gross indicator of the positive charging of grains in general, which is generally quantified by the charging parameter ($\gamma$) as shown in Sect.~\ref{subsec:pe_gamma}.  Figure~\ref{fig:correlation} directly compares of the observationally derived \pe\ and $f(+)$, independent of the use of PDR models to quantify the physical conditions, over a wide range of the physical properties of PDRs.

Using (\oi\ 63\um\ $+$ \oi\ 145\um\ $+$ \cii\ 158\um) / TIR as a tracer of \pe\ contains several assumptions.  We neglect the heating by the collisional de-excitation of vibrationally excited \hh.  The \hh\ de-excitation heating becomes a dominant heating process in a dense PDR with a low UV field \citep{Roellig2006}.  Among our PDRs, only $G_0$ and the upper value of the density of the Horsehead are close to the regime where the contribution from the photoelectric heating and \hh\ heating is comparable.  Those in other PDRs indicate that the photoelectric heating is dominant.  We also neglect the cooling by \hh\ emission.  This is partly justified because the \hh\ emissions as a consequence of the excitation by the UV-pumping and the \hh\ formation do not contribute to the estimate of the photoelectric heating efficiency.  \citet{Habart2011} showed intense \hh\ lines in the Horsehead (0--0 S(0) to S(3) and 1--0 S(1)), which are, as a sum, comparable with the (\oi\ $+$ \cii) intensity.  In NGC~7023~E and Ced~201, the sum of the \hh\ pure rotational emissions S(0) to S(3) (\citeauthor{Habart2011}~\citeyear{Habart2011} for NGC~7023~E, and from the IRS spectra in the \textit{Spitzer} archive for Ced~201) is $\sim 80$\% and $\sim 60$\% of the (\oi\ $+$ \cii) intensity, respectively.  Neglecting these \hh\ emissions may underestimate \pe\ by a factor of $\lesssim 2$.  In other regions, the \hh\ pure rotational emissions from the IRS spectra in the \textit{Spitzer} archive show the intensity of $\lesssim 20$\% of (\oi\ $+$ \cii).

As mentioned in Sect.~\ref{subsec:reduction_herschel}, the \oi\ 63\um\ emission is indicated to be optically thick in most targets.  Therefore, the optical-depth-corrected \oi\ 63\um\ intensity is higher than the observed values.  However, we observe the emission that leaves the cloud, and this contributes to the net cooling of the cloud.  This justifies the direct use of the observed (\oi\ $+$ \cii) to express the gas cooling, except if the foreground absorption is significant, which is unlikely for our PDRs, as mentioned above.

As discussed for $G_0$(TIR) (Sect.~\ref{subsec:analysis_fir}), photons with an energy of $<6$~eV contribute to TIR, but not to the photoelectric heating.  Therefore, estimating \pe\ with the observed TIR provides a lower limit.  For the exciting source(s) of spectral types earlier than early B, the effect is not significant in terms of the absolute factor and the variation between different spectral types.  For Ced~201, which has an exciting source of B9.5, we may underestimate \pe\ by a factor of a few.  This uncertainty, however, does not change the overall trend in Fig.~\ref{fig:correlation} and our conclusions.

Our definition of \pe\ assumes a spherical geometry, where both line and continuum emissions radiate isotropically.  If one assumes an edge-on geometry, where the continuum emission can escape through the molecular side while the line emissions, especially \oi\ 63\um, becomes completely optically thick and can be emitted only from the front side, \pe\ should be divided by 2 \citep{Tielens2005}.

\subsection{Comparison of \pe\ with theory}\label{subsec:pe_gamma}

\begin{figure}
\centering
\includegraphics[width=0.49\textwidth]{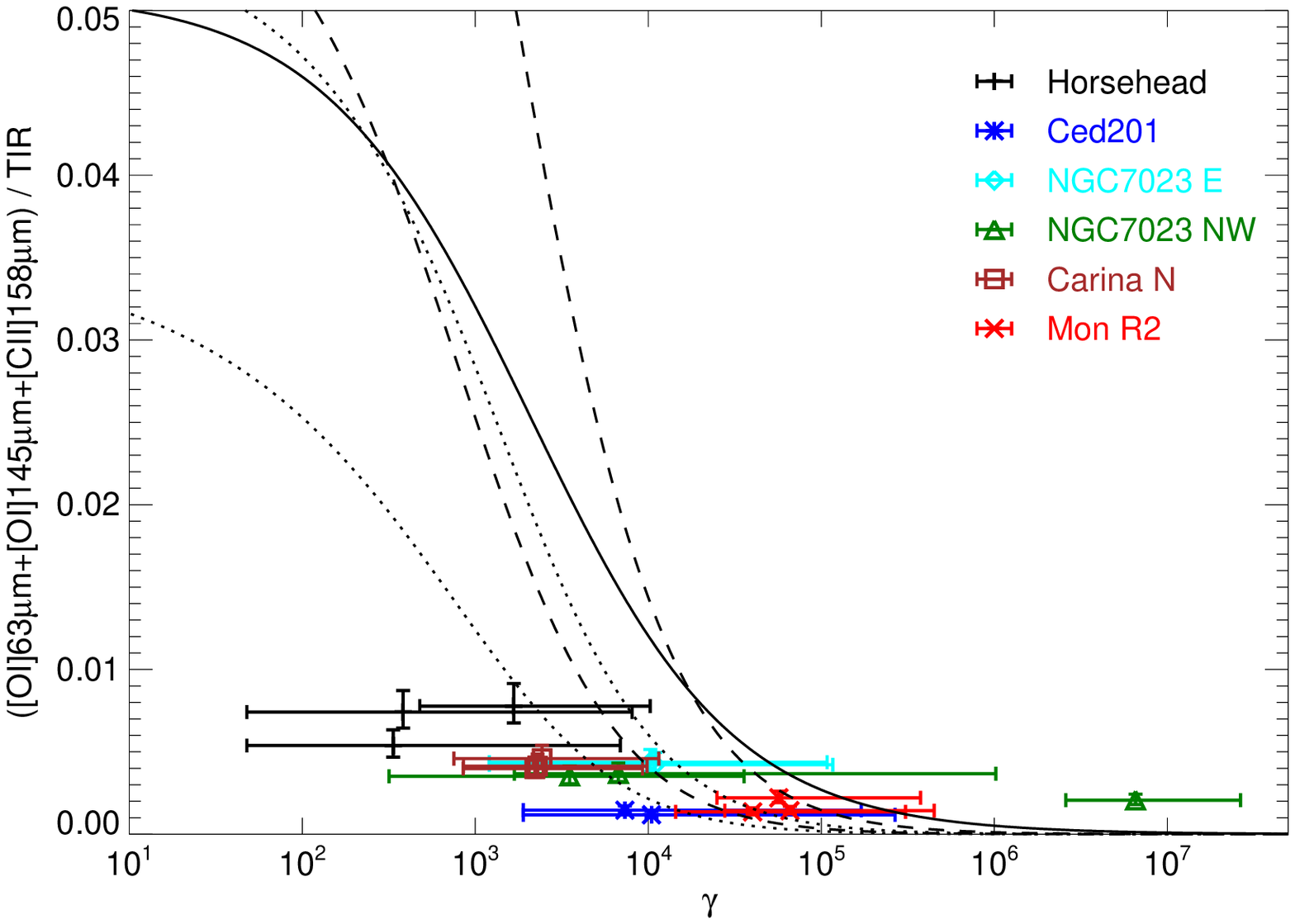}
\caption{\pe $=$ (\oi\ 63\um\ $+$ \oi\ 145\um\ $+$ \cii\ 158\um) / TIR versus the charging parameter ($\gamma=G_0 T^{1/2}/n_e$).  Solid line is the theoretical calculation from \citet{Bakes1994}, dashed lines are that from \citet{WD2001pe} for $R_V=3.1$ ($10^5b_\mathrm{C}=0.0$ and $6.0$ for the lower and upper line), dotted lines are for $R_V=5.5$ with Case A ($10^5b_\mathrm{C}=0.0$ and $3.0$ for the lower and upper line; see text).}
\label{fig:tir_gamma}
\end{figure}

\begin{figure}
\centering
\includegraphics[width=0.49\textwidth]{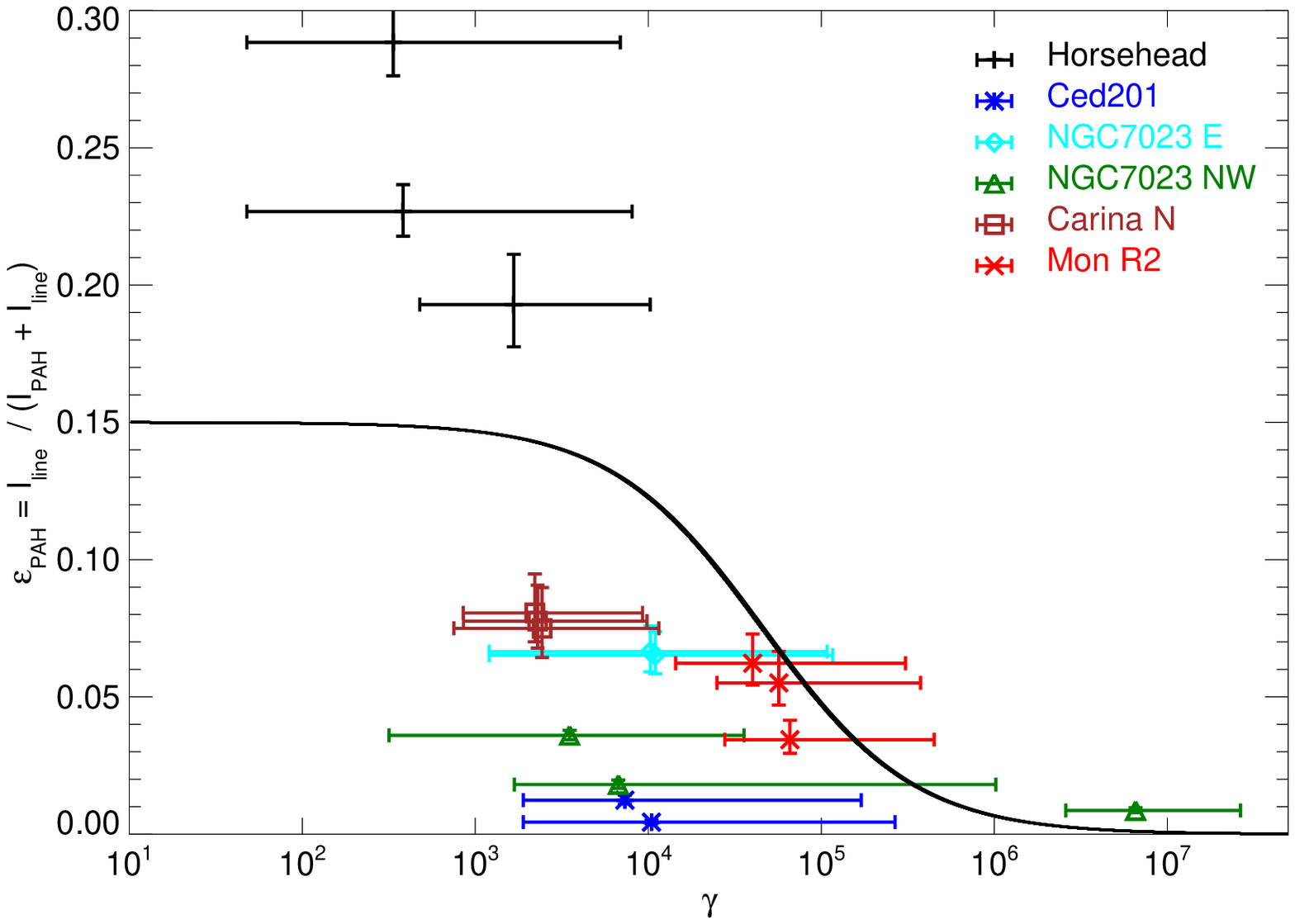}
\caption{\pepah $=$ (\oi\ 63\um\ $+$ \oi\ 145\um\ $+$ \cii\ 158\um) / (PAH $+$ \oi\ 63\um\ $+$ \oi\ 145\um\ $+$ \cii\ 158\um) versus the charging parameter ($\gamma$).  The black curve depicts our calculations using 25 different size distributions from \citet{WD2001}, which looks degenerated (see text).}
\label{fig:band_gamma}
\end{figure}

\begin{figure}
\centering
\includegraphics[width=0.49\textwidth]{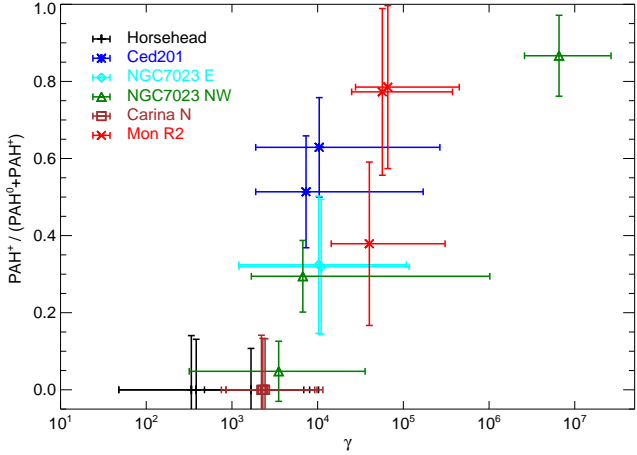}
\caption{Fraction of \pahp\ ($f(+)$) versus the charging parameter ($\gamma$).}
\label{fig:pahp_gamma}
\end{figure}

Theoretically, \pe\ can be expressed as a function of the charging parameter, defined as $\gamma=G_0 T^{1/2}/n_e$, where $T$ is the gas temperature and $n_e$ is the electron density.  $\gamma$ is proportional to the ratio of the ionization and recombination rate \citep{Bakes1994}.  We calculated $\gamma$ from $G_0$, \nh, and $T$ listed in Table~\ref{table:tir_and_uv}.  They were carefully estimated in each subregion (Appendix~\ref{sect:app_gamma}) to be independent of the existing PDR models.  We estimated $n_e$ by assuming that most electrons are provided by carbon ionization and all carbon atoms are singly ionized, i.e., $n_e=x(\mathrm{C})$\nh, where $x(\mathrm{C})=1.6\times 10^{-4}$ \citep{Sofia2004} is the elemental carbon abundance.  In Fig.~\ref{fig:tir_gamma} we show the relation between \pe\ and $\gamma$.  There is a trend that \pe\ decreases when $\gamma$ increases.  \citet{Bakes1994} theoretically estimated \pe\ based on the physics of the photoelectric effect on dust grains.  \citet{WD2001pe} modeled \pe\ with some improved physical parameters and various size distributions of dust grains in \citet{WD2001}.  Dashed and solid lines in Fig.~\ref{fig:tir_gamma} show some of their estimates, representing the strongest variation of \pe\ depending on the dust size distribution.  Dense regions characterized by $R_v=5.5$ (dotted lines) have a lower heating efficiency.  The parameter $b_\mathrm{C}$ is the total C abundance in the log-normal populations of the size distribution, and the model with a larger $b_\mathrm{C}$ shows a stronger enhancement at the size of $<10$\AA\ \citep{WD2001}.  Case A is the case without the constraint on the total grain volume \citep{WD2001}.  Although the observed trend that \pe\ decreases when $\gamma$ increases is reproduced by models, the observation indicates a weaker dependence of \pe\ on $\gamma$, and all models overestimate \pe\ at low $\gamma$.  In the following, we focus on the photoelectric heating only on PAHs.

When we consider \pe\ on PAHs, (\oi\ 63\um\ $+$ \oi\ 145\um\ $+$ \cii\ 158\um) / TIR is not an appropriate definition, because the TIR expresses the total energy emitted by all dust grains.  The total energy that PAHs absorb is converted into either MIR AIB emission or cooling line emission through the photoelectric effect on PAHs.  A fraction of the energy (generally taken to be $0.5$) remains behind as electronic excitation energy after the photoelectric effect \citep{Tielens2005}, which also results in MIR AIB emission.  Therefore \pepah\ is defined as (\oi\ 63\um\ $+$ \oi\ 145\um\ $+$ \cii\ 158\um) / (PAH band emission $+$ \oi\ 63\um\ $+$ \oi\ 145\um\ $+$ \cii\ 158\um).  The trend of \pepah\ against $\gamma$ (Fig.~\ref{fig:band_gamma}) is very similar to that reported in Fig.~\ref{fig:tir_gamma}.  

Following \citet{Tielens2005}, we computed a simple theoretical estimate of \pepah.  For a given PAH containing the number of carbon atoms of $N_c$,
\begin{equation}
\epsilon_{\mathrm{PAH},N_c}=\frac{1}{2}f(Z=0)\left(\frac{h\nu-IP}{h\nu}\right), \label{eq:pepah_ori}
\end{equation}
where $f(Z=0)$ is the neutral fraction and $IP$ is the ionization potential.  With a typical photon energy of 10~eV and an ionization potential of 7~eV, this equation becomes
\begin{equation}
\epsilon_{\mathrm{PAH},N_c}=0.15f(Z=0). \label{eq:pepah}
\end{equation}
The neutral fraction is given by
\begin{equation}
f(Z=0)=(1+\gamma_0)^{-1}, \label{eq:neutralfr}
\end{equation}
where $\gamma_0$ is the ratio of the ionization rate over the recombination rate, which for small PAHs is given by
\begin{equation}
\gamma_0=3.5\times 10^{-6}N_c^{1/2}\gamma.  \label{eq:gamma}
\end{equation}
We substitute Eq.(\ref{eq:gamma}) into Eq.(\ref{eq:neutralfr}) and then into Eq.(\ref{eq:pepah}), and integrate over the size distribution from \citet{WD2001} up to $N_c=100$.  In contrast to \pe, in which a different size distribution gives a large difference (Fig.~\ref{fig:tir_gamma}), \pepah\ is insensitive to the adopted size distribution.  In Fig.~\ref{fig:band_gamma}, the results with 25 different size distributions in \citet{WD2001} are plotted, but they look degenerated.  Although the match between the theoretical estimates and the observed values is not excellent, it is much better than in Fig.~\ref{fig:tir_gamma}.  This result suggests that the photoelectric heating is dominated by PAHs.

In Fig.~\ref{fig:pahp_gamma}, the relation between the fraction of \pahp\ ($f(+)$) and $\gamma$ is shown.  PAHs are almost neutral at $\gamma<10^3$ and almost fully ionized at $\gamma>10^6$, and there is a transition in-between.  Since $\gamma$ is defined by the environment properties and it is proportional to the ratio of the grain ionization and recombination rate, Fig.~\ref{fig:pahp_gamma} confirms that the fraction of \pahp\ is a good indicator for the positive charging of grains in general.  This trend is also consistent with the correlation between the intensity ratio of PAH(6.2\um)/PAH(11.3\um) and $\gamma$ presented in \citet{Galliano2008}.


\section{Summary}

We analyzed \textit{Herschel}/PACS and \textit{Spitzer}/IRS spectroscopic observations in six PDRs and showed that the photoelectric heating efficiency (\pe) is lower in regions with a large fraction of positively ionized PAHs ($f(+)$).  Based on examining the photoelectric heating efficiency on PAHs, we found a dominant contribution of PAHs to the photoelectric heating.

Our PDR sample covers a wide range of physical conditions ($100\lesssim G_0 \lesssim 10^5$) and provides a good test case for investigating the relation between \pe\ and the charge state of PAHs.  We estimated \pe\ as (\oi\ 63\um\ $+$ \oi\ 145\um\ $+$ \cii\ 158\um) / TIR.  \pe\ varies in our PDRs between 0.1 and 0.9\%.  $f(+)$ was obtained from a fit of the MIR spectra with a set of template spectra representing PAH-related species and varied from 0\% ($+11$\%) to 87\% ($\pm 10$\%).  All positions with a high $f(+)$ show a low \pe, and all positions with a high \pe\ show a low $f(+)$.  This trend supports a scenario in which a positive grain charge results in a decreased heating efficiency.  The theoretical estimate of \pe\ shows a stronger dependence on the charging parameter ($\gamma$) than the observed \pe\ reported in this study, and overestimates \pe\ at low $\gamma$.  The photoelectric heating efficiency on PAHs, \pepah $=$ (\oi\ 63\um\ $+$ \oi\ 145\um\ $+$ \cii\ 158\um) / (PAH band emission $+$ \oi\ 63\um\ $+$ \oi\ 145\um\ $+$ \cii\ 158\um), shows a much better match between the observations and the theoretical estimates, indicating a dominant contribution of PAHs on the photoelectric heating.  PDR models that fully account for the relative contribution of different PAH and eVSGs populations are needed.  Velocity-resolved observations of \oi\ 63\um\ in the future with for instance SOFIA/upGREAT \citep{Heyminck2012} will enable us to investigate the foreground absorption and the optical depth effect in detail \citep[c.f.][]{Boreiko1996}.

\begin{acknowledgements}
PACS has been developed by a consortium of institutes led by MPE (Germany) and including UVIE (Austria); KU Leuven, CSL, IMEC (Belgium); CEA, LAM (France); MPIA (Germany); INAF-IFSI/OAA/OAP/OAT, LENS, SISSA (Italy); IAC (Spain). This development has been supported by the funding agencies BMVIT (Austria), ESA-PRODEX (Belgium), CEA/CNES (France), DLR (Germany), ASI/INAF (Italy), and CICYT/MCYT (Spain).  HIFI has been designed and built by a consortium of institutes and university departments from across Europe, Canada and the United States under the leadership of SRON Netherlands Institute for Space Research, Groningen, The Netherlands and with major contributions from Germany, France and the US. Consortium members are: Canada: CSA, U.Waterloo; France: CESR, LAB, LERMA, IRAM; Germany: KOSMA, MPIfR, MPS; Ireland, NUI Maynooth; Italy: ASI, IFSI-INAF, Osservatorio Astrofisico di Arcetri-INAF; Netherlands: SRON, TUD; Poland: CAMK, CBK; Spain: Observatorio Astron\'{o}mico Nacional (IGN), Centro de Astrobiolog\'{i}a (CSIC-INTA). Sweden: Chalmers University of Technology - MC2, RSS \& GARD; Onsala Space Observatory; Swedish National Space Board, Stockholm University - Stockholm Observatory; Switzerland: ETH Zurich, FHNW; USA: Caltech, JPL, NHSC.

We thank the \textit{Herschel} helpdesk and the \textit{Spitzer} helpdesk for their support in analyzing data.  We thank the referee for useful suggestions that greatly improved the paper.  Part of this work was supported by the German \emph{Deut\-sche For\-schungs\-ge\-mein\-schaft, DFG\/}, project number SFB956 C1, by the Spanish program CONSOLIDER INGENIO 2010, under grant CSD2009-00038 Molecular Astrophysics: The Herschel and ALMA Era (ASTROMOL), by CNES, by a Ram\'on y Cajal research contract, and by the Spanish MICINN through grants AYA2009-07304 and CSD2009-00038.

\end{acknowledgements}

\bibliographystyle{aa}

\begin{appendix}
\section{Relative flux uncertainty between 63\um\ and 158\um\ caused by the difference of the beam size}
\label{sect:app_beamsize}

\begin{figure}
\centering
\includegraphics[width=0.49\textwidth]{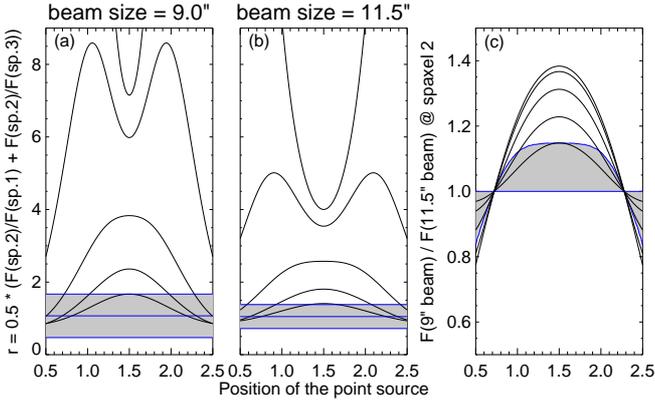}
\caption{Black curves show the simulated $r$, defined as Eq.~(\ref{eq:r}), as a function of the position of the point source for the beam size of (a) 9\arcsec\ and (b) 11.5\arcsec.  The relative contribution of the extended source compared to the height of the point source at 63\um\ is $0.0$, $0.02$, $0.1$, $0.3$, and $0.7$ from upper to lower curves.  Shaded areas represent the observed values with errors of $\pm 3\sigma$ (blue lines in Fig.~\ref{fig:beamdiff_obs}). (c) The flux ratio of 9\arcsec\ beam and 11.5\arcsec\ beam at spaxel 2 as a function of the position of the point source with the same model.  Shaded areas show ranges when $r$ in (a) or (b) matches the observed value.}
\label{fig:beamdiff_sim}
\end{figure}

\begin{figure}
\centering
\includegraphics[width=0.49\textwidth]{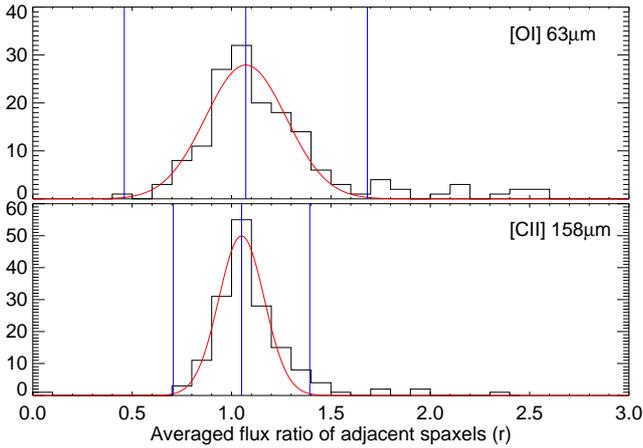}
\caption{Histograms of the observed $r$ for \oi\ 63\um\ and \cii\ 158\um.  Red lines show the fit with a Gaussian profile, and blue lines show the center and $\pm 3\sigma$ of the fitted Gaussian.}
\label{fig:beamdiff_obs}
\end{figure}

We estimate the uncertainty of the relative flux in one spaxel at different wavelengths that arises from the different PSF and the unknown spatial structure of the sources by considering the combination of two extremes, pure point sources and flat extended emission, and by comparing the contrast between neighboring spaxels for these combinations with the contrast actually measured in the observations.

We consider three spaxels 1--3 along a line whose centers are located at ($0.5$,$0$), ($1.5$,$0$), and ($2.5$,$0$) in units of a spaxel size ($9.4$\arcsec).  We put a point source represented by a Gaussian profile with the FWHM of 9\arcsec\ and 11.5\arcsec\ for 63\um\ and 158\um, respectively, centered at various positions between ($0.5$,$0$) and ($2.5$,$0$), and calculate the mean flux ratio of the adjacent spaxels as
\begin{equation}
r=\frac{1}{2}\left(\frac{F(\mathrm{spaxel\ 2})}{F(\mathrm{spaxel\ 1})}+\frac{F(\mathrm{spaxel\ 2})}{F(\mathrm{spaxel\ 3})}\right), \label{eq:r}
\end{equation}
where $F$ is the flux falling in one spaxel. $r$ is lowest when the center of the point source is located at ($1.5$,$0$), which is the center of spaxel 2, and the minimum $r$ is $7.1$ and $4.0$ for 63\um\ and 158\um, respectively (upper lines in Figs.~\ref{fig:beamdiff_sim}a,b).  On the other hand, we derive the observed $r$ of the \oi\ 63\um\ and \cii\ 158\um\ line intensities from the original $5\times 5$ PACS spaxels for all regions.  The histogram of the distribution of $r$ is well represented by a Gaussian with a small number of outliers at high $r$ values (Fig.~\ref{fig:beamdiff_obs}), and the fitted Gaussian has a center of $1.1$ for both lines and $\sigma$ of $0.2$ and $0.1$ for 63\um\ and 158\um.  Even the upper $3\sigma$ ($1.7$ and $1.4$) is smaller compared to the above model with a point source, indicating that the contribution of extended emission is significant in our regions.  Therefore, we modify the model by adding a constant extended component to all spaxels, and estimate $r$, which is shown as black lines in Figs.~\ref{fig:beamdiff_sim}a and b with varying constant values.  Then we estimate F(spaxel~2, 63\um)/F(spaxel~2, 158\um), which expresses the uncertainty of the relative flux at 63\um\ and 158\um, for each model (black lines in Fig.~\ref{fig:beamdiff_sim}c).  The observed $r$ with $3\sigma$ errors are shown as shaded areas in Figs.~\ref{fig:beamdiff_sim}a and b, and the corresponding ranges are shaded in Fig.~\ref{fig:beamdiff_sim}c, which is numerically calculated by surveying the constant emission value and determining the position range of the point source, where $r$ is in the observed range for each constant emission value.  It reaches to within 15\% of unity.  Therefore, we take 15\%\ as the uncertainty of the relative flux between 63\um\ and 158\um\ caused by the beam size difference.

\section{Estimate of the charging parameter}
\label{sect:app_gamma}

Here we describe the estimate of $G_0$, the gas density \nh, and the gas temperature $T$ to be used for deriving the charging parameter $\gamma$ in individual subregions (Table~\ref{table:tir_and_uv}).  The rotational temperature derived from the low-$J$ \hh\ pure rotational emissions up to S(4) is adopted as $T$ in all regions.  $G_0$ and \nh\ affect $\gamma$ linearly and we estimate them independent of an existing PDR model.  These properties are representative of the warm PDR surface where most of the \hh\ rotational emission and PAH emission comes from.

\subsection{Horsehead}
\citet{Habart2005} estimated $G_0$(star) $\sim 100$ at the PDR interface based on the effective temperature of the exciting star and assuming geometrical dilution at the projected distance. Under this assumption, our three subregions are close enough to each other to make no difference for the $G_0$(star) between them.  $G_0$ derived from TIR, $G_0$(TIR), is listed in Table~\ref{table:tir_and_uv}.  We adopt both $G_0$(star) and $G_0$(TIR) for the possible range in individual subregions.  For the gas temperature $T$, we use the range of the estimate from the intensity ratio of \hh\ S(2)/S(0), S(3)/S(1), and S(4)/S(2) in \citet{Habart2011}.

\citet{Abergel2003} showed that the lower limit of the density behind the filament is $\sim 2\times 10^4$~\cc\ from the infrared brightness profile.  \citet{Habart2005} modeled the spatial distribution of \hh, PAH, CO and 1~mm continuum emissions using a PDR model and suggested the density gradient from \nh$=10^4$~\cc\ in the \hh\ emitting region to \nh$=2\times 10^5$~\cc\ in the inner cold molecular layers.  Since the region we are interested in is the \hh-emitting gas, we still include the \nh\ value in the \hh-emitting region even for subregion 3, i.e. we adopt \nh$=$($0.1$--$2$)$\times 10^5$~\cc\ for subregions 2 and 3.  \nh$\sim 10^4$~\cc\ at the interface is also confirmed by the pressure equilibrium with the ionized gas.  In the ionized gas, $n_e$ is estimated to be $100$--$350$~\cc\ with $T_e=7500$~K \citep{Compiegne2007}.  Assuming the pressure equilibrium, $2n_eT_e=$\nh$T$, gives \nh $=$($0.3$--$2$)$\times 10^4$~\cc.  Therefore, we adopt \nh$=$($0.3$--$2$)$\times 10^4$~\cc\ for subregion 1.

\subsection{Ced~201}

\citet{Kemper1999} estimated $G_0=200$ by comparing the observations with a PDR model in units of the average interstellar radiation between 2~eV and 13.6~eV.  On the other hand, \citet{YoungOwl2002} derived $G_0=300$ by the infrared continuum emission, with a geometry correction factor of 2.  We adopt $G_0=200$ as a lower limit and $G_0$(FIR) as an upper limit for each subregion.  The fraction of eVSGs determined by PAHTAT is correlated to $G_0$ \citep{Pilleri2012a} and the adopted $G_0$ for each subregion is also consistent with it.

For \nh, \citet{YoungOwl2002} derived \nh $=4\times 10^2$~\cc\ from the lower limit of \cii\ 158\um/\oi\ 63\um\ using a PDR model.  \citet{Kemper1999} estimated \nh\ of $(5\pm 1)\times 10^3$~\cc\ by simple excitation models of CO and $^{13}$CO emissions.  They also modeled the emission of CO, C, C$^+$, CS, and HCO$^+$ using a PDR model, which suggests \nh\ of $1.2\times 10^4$~\cc.  We use all these ranges for the two subregions.  $T$ is estimated as $\sim 330$~K from the ratio of \hh\ S(1) and S(3) emissions \citep{Kemper1999}.

\subsection{NGC~7023~E}

\citet{Pilleri2012a} modeled the spatial distribution of the MIR AIB emission, assuming a spherical shell geometry and the energy balance taking into account, and derived the spatial variation of \nh\ and $G_0$ along a cut in NGC~7023~E.  At the PDR front, $G_0$ is calculated to be 250 based on the spectral type of the star, geometrical dilution and assuming an attenuation of $A_v=1.5$ around the star \citep{Pilleri2012a}.  The density at the PDR front is derived as $1.4\times 10^3$~\cc.  Around the MIR AIB emission peak, which corresponds to the regions in this study, $G_0$ is estimated to be $120$--$170$ and \nh $=(0.5$--$1)\times 10^4$~\cc.  On the other hand, $G_0$(TIR) is 970 and 900 for two subregions.  We adopt 120 as a lower limit of $G_0$, and the corresponding $G_0$(TIR) as an upper limit for each subregion, and \nh $=(0.1$--$1)\times 10^4$~\cc\ for both subregions.  For the gas temperature $T$, we use $258$--$370$~K from the intensity ratio of \hh\ S(2)/S(0) and S(3)/S(1) \citep{Habart2011}.

\subsection{NGC~7023~NW}

The same modeling as for NGC~7023~E was made by \citet{Pilleri2012a}.  Our three subregions are not exactly on their cut because we chose the subregions to maximize the variation of the \pahp\ fraction.  Subregion 2 is located at the PDR interface.  \citet{Pilleri2012a} estimate $G_0$(star)$=2600$ at the PDR front, whereas $G_0$(TIR) $=7700$ at this subregion.  We adopt $G_0=2600$--$7700$ for subregion 2.  Subregion 1 is located closer to the star than the PDR interface.  The scaling of $G_0$(star) by the projected distance gives $G_0=10^4$, higher than $G_0$(TIR), possibly because either the real distance to the star is larger than the projected distance, and/or the assumption that all UV radiation is absorbed and converted into the IR emission underestimates $G_0$ in such a cavity because many UV photons pass the region.  Nevertheless, we conservatively cover both values, i.e., $G_0=(0.5$--$1)\times 10^4$, for subregion 1.  The adopted $G_0$ in subregions 1 and 2 is also consistent with the correlation between the fraction of eVSGs and $G_0$ from \citet{Pilleri2012a}.  For subregion 3, we use this correlation to derive $G_0$ \citep[see Eq.(5) in][]{Pilleri2012a}, which results in $G_0=500$--$3100$.  Together with $G_0$(TIR)$=5400$, we adopt $G_0=500$--$5400$ for subregion 3.

In \citet{Pilleri2012a}, the modeled density profile sharply increases from $1.1\times 10^3$~\cc\ to $2\times 10^4$~\cc\ within $\sim 8$\arcsec.  \citet{Fuente1996} showed high-density filaments of a few $10^5$~\cc\ based on HCO$^+$ observations.  Therefore, we adopt $10^3$--$2\times 10^5$~\cc\ for subregion 2, and ($0.2$--$2$)$\times 10^5$~\cc\ for subregion 3.  \citet{Berne2012} examined \nh\ in the cavity in detail from several diagnostics and suggested $150\pm 100$~\cc.  We take this as \nh\ for our subregion 1.  For the gas temperature $T$, we use $430$--$450$~K from the intensity ratio of \hh\ S(3)/S(1) and S(4)/S(2) \citep{Fuente2000}.

\subsection{Carina~N}

We estimate $G_0$(star) by computing the contribution from all OB stars of Trumpler 14 listed in \citet{Smith2006}, which gives ($7$--$8$)$\times 10^3$ for the three subregions.  On the other hand, $G_0$(TIR) is ($1.5$--$1.6$)$\times 10^4$.  The point-source catalog of the Wide Field Infrared Survey Explorer \citep[WISE;][]{Wright2010} contains several sources with a flux stronger at 4.6\um\ than at 3.4\um\ close to our region, which indicates a contribution from embedded protostars.  We consider both $G_0$(star) and $G_0$(TIR) as a possible range in individual subregions.

\citet{Kramer2008} modeled CO and \ci\ emissions in the Carina~N region and suggested \nh\ of $2\times 10^5$~\cc.  This is consistent with a pressure equilibrium with the \hii\ region.  Using the emission lines ratio from ions with similar ionization potential observed with IRS, we can estimate $n_e$ assuming the elemental solar abundance and isothermal thin emission.  \neii\ 12.8\um/\suiii\ 18.7\um, \piii\ 17.9\um/\neii\ 12.8\um, and \ariii\ 8.99\um/\suiii\ 18.7\um\ give $n_e$ of $10^4$, $2\times 10^4$, and $9\times 10^3$--$10^4$~\cc, respectively, when $T_e=10^4$~K.  Since the gas temperature $T$ in PDRs is derived as $T=290$--$529$~K from \hh\ emission lines, \nh\ can be estimated as $3\times 10^5$--$10^6$~\cc.  We use $2\times 10^5$--$10^6$~\cc\ for all subregions.

\subsection{Mon~R2}

We calculate $G_0$(star) from the contribution of IRS1-4 (see Fig.~\ref{fig:obspos}) with the luminosity listed in \citet{Henning1992}, as ($5.2$--$5.9$)$\times 10^4$, $10^5$, and ($1.1$--$1.2$)$\times 10^5$ in the three subregions.  While $G_0=5\times 10^5$ is used to characterize the radiation field at the ionization front \citep{Rizzo2003}, our subregions do not cover the nearest regions of infrared sources to avoid unresolved complex spatial distributions.  $G_0$(star) matches $G_0$(TIR) well.  We consider both $G_0$(star) and $G_0$(TIR) as a possible range in individual subregions.

\citet{Rizzo2005} and \citet{Pilleri2013} performed an LVG analysis with C$_3$H$_2$ and C$_2$H at several positions in Mon~R2 and derived \nh\ $>10^6$~\cc.  \citet{Ginard2012} showed similar results from different molecules; the molecular hydrogen density of a few $10^5$~\cc\ to $10^6$~\cc.  These molecules traces the cold molecular gas with a temperature of $T\sim 45$~K \citep{Giannakopoulou1997}.  On the other hand, \citet{Berne2009} derived \nh\ of ($0.4$--$4$)$\times 10^5$ from PDR model calculations using \hh\ rotational emission lines, and the temperature derived from the \hh\ rotational emission lines is $T=314$--$574$~K.  These \nh\ and $T$ are consistent with a pressure equilibrium with the cold molecular gas.  In the \hii\ region, the emission measure of $1.9\times 10^7$~pc\,cm$^{-6}$ and the geometrical mean diameter of 23.7\arcsec\ \citep{Takahashi2000,Wood1989} give an estimate of $n_e=1.4\times 10^4$~\cc.  With $T_e=7600$~K \citep{Downes1975} and $T=314$--$574$~K, the pressure equilibrium gives \nh\ $=$($4$--$7$) $\times 10^5$~\cc, which is also compatible with the estimate from \hh\ rotational emissions, although the pressure equilibrium is an inadequate assumption in an UC\hii\ like Mon~R2.  We adopt the estimate range from \hh\ emissions, i.e., ($0.4$--$4$)$\times 10^5$ for all subregions.

\end{appendix}
\end{document}